\documentclass[aps,prl,superscriptaddress,floatfix,reprint]{revtex4-1}  

\usepackage{graphicx}
\usepackage{bm}
\usepackage{amssymb}
\usepackage{amsmath}
\usepackage{float}
\usepackage{pstricks} 
\usepackage{color}
\usepackage{bbold}
\usepackage{xr} 

\newcommand{\bra}[1]{\left\langle #1 \right\vert}
\newcommand{\ket}[1]{\left\vert #1 \right\rangle}
\newcommand{\braket}[2]{\left\langle #1 \middle\vert #2 \right\rangle}
\newcommand{\proj}[3]{\ket{#1}_{\!#2}\!\!\bra{#3}}

\begin{document}

\title{A chip-based array of near-identical, pure, heralded single photon sources}

\author{Justin~B.~Spring}
\affiliation{Clarendon Laboratory, University of Oxford, Parks Road, Oxford OX1 3PU, UK}

\author{Paolo~L.~Mennea}
\affiliation{Optoelectronics Research Centre, University of Southampton, Southampton, SO17 1BJ, UK}

\author{Benjamin~J.~Metcalf}
\affiliation{Clarendon Laboratory, University of Oxford, Parks Road, Oxford OX1 3PU, UK}

\author{Peter~C.~Humphreys}
\affiliation{Clarendon Laboratory, University of Oxford, Parks Road, Oxford OX1 3PU, UK}

\author{James~C.~Gates}
\affiliation{Optoelectronics Research Centre, University of Southampton, Southampton, SO17 1BJ, UK}

\author{Helen~L.~Rogers}
\affiliation{Optoelectronics Research Centre, University of Southampton, Southampton, SO17 1BJ, UK}

\author{Christoph~S\"{o}ller}
\affiliation{Clarendon Laboratory, University of Oxford, Parks Road, Oxford OX1 3PU, UK}

\author{Brian~J.~Smith}
\affiliation{Clarendon Laboratory, University of Oxford, Parks Road, Oxford OX1 3PU, UK}

\author{W.~Steven~Kolthammer}
\affiliation{Clarendon Laboratory, University of Oxford, Parks Road, Oxford OX1 3PU, UK}

\author{Peter~G.~R.~Smith}
\affiliation{Optoelectronics Research Centre, University of Southampton, Southampton, SO17 1BJ, UK}

\author{Ian~A.~Walmsley}
\affiliation{Clarendon Laboratory, University of Oxford, Parks Road, Oxford OX1 3PU, UK}
\email{ian.walmsley@physics.ox.ac.uk}

\date{\today}

\begin{abstract}{Interference between independent single photons is perhaps the most fundamental interaction in quantum optics. It has become increasingly important as a tool for optical quantum information science, as one of the rudimentary quantum operations, together with photon detection, for generating entanglement between non-interacting particles. Despite this, demonstrations of large-scale photonic networks involving more than two independent sources of quantum light have been limited due to the difficulty in constructing large arrays of high-quality single photon sources. Here, we solve the key challenge, reporting a novel array of more than eighteen near-identical, low-loss, high-purity, heralded single photon sources achieved using spontaneous four-wave mixing (SFWM) on a silica chip. We verify source quality through a series of heralded Hong-Ou-Mandel experiments, and further report the experimental three-photon extension of the entire Hong-Ou-Mandel interference curves, which map out the interference landscape between three independent single photon sources for the first time.}
\end{abstract}

\maketitle

Quantum states of many particles offer an opportunity to study the rich physics of large-scale quantum correlations. Light provides the possibility of building such quantum states under ambient conditions, both because photons are robust to dephasing due to minimal interaction with their environment and one another, and because entanglement can be generated between photons allowing the observation of strong quantum phenomena in everyday settings~\cite{URen2003, Zukowski1995}. Consequently, quantum optical networks are expected to enable new technologies - spanning communications, distributed sensing, simulation, and computation  - characterized by performance that dramatically exceeds their classical counterparts~\cite{Walmsley2005a}.

Entanglement can be established across such a quantum network through quantum interference of independent single photons and measurement using a photodetector~\cite{Knill2001}.  The archetype of this interaction is the Hong-Ou-Mandel (HOM) effect~\cite{Intervals1987}, whereby two photons from independent sources coalesce at a beam splitter, always leaving at the same output. Remarkably, this approach enables the generation of entanglement directly from the bosonic properties of the underlying fields, without the necessity for the photons to interact directly. 

A key challenge is scaling this approach to generate the large entangled networks necessary in many technological applications. Over the past decade, rapid progress has been made in developing several of the elements necessary to construct such large-scale quantum optical states. High efficiency photon-counting detectors~\cite{Calkins2013}, actively programmable circuits~\cite{Carolan2015,Metcalf2014qtp}, and quantum memories have all been demonstrated~\cite{Heshami2015}. A significant remaining obstacle is the demonstration of arrays of many quantum light sources that deliver single photons with low loss, high purity, and sufficient control of the full photonic modal structure that allows high quality interference. 

The main approaches to preparing single photons in such states are either to place a single emitter in a high-finesse cavity or to use a nonlinear optical process by which photons are produced in pairs by spontaneous scattering, and the detection of one heralds the presence of the other~\cite{Eisaman2011}. Both approaches offer routes to deterministic photon generation, though the latter requires effective routing of multiple pair sources~\cite{Migdall2002}. Despite this complication, heralded photon-pair sources have become the de-facto standard for generating quantum states of light for networking by virtue of their ease of use and operation in ordinary laboratory conditions. 

\begin{figure*}[ht]
\begin{center}
\includegraphics[width=\textwidth]{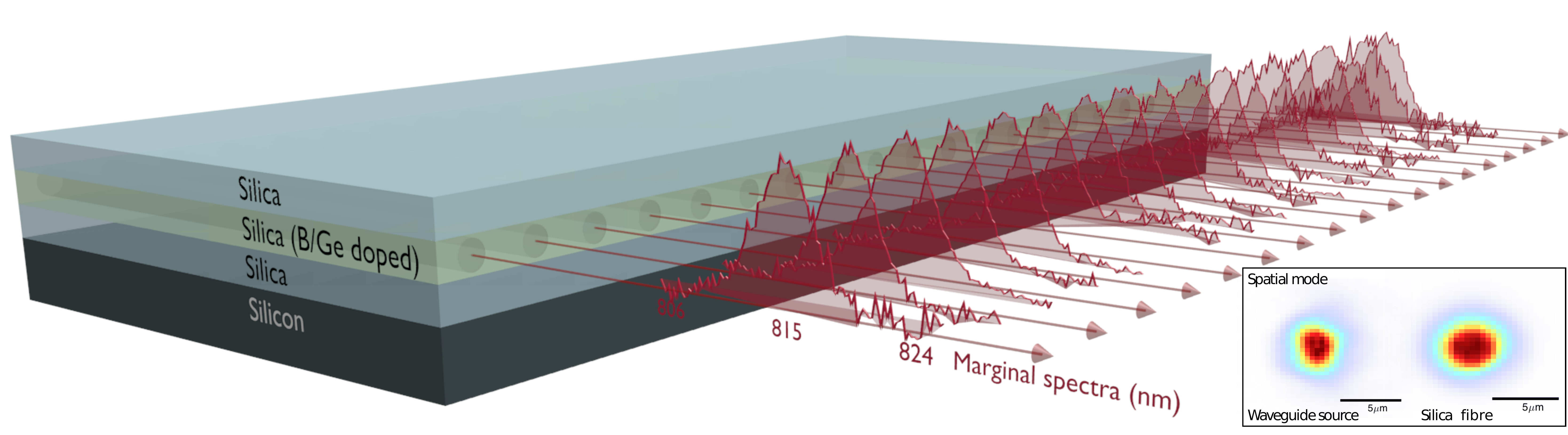}
\caption{{\bf An array of heralded single photon sources on a silica photonic chip}. A series of straight waveguides are fabricated via UV-laser writing in a germanium-doped silica-on-silicon photonic chip, each of which constitutes its own heralded single photon source. Spontaneous four-wave mixing (SFWM) is achieved through birefringent phase matching and generates correlated pairs of photons. Each source is pumped using the same pulsed laser centred at 736\,nm and the marginal spectrum of each heralded photon is recorded using a single-photon spectrometer. We find a mean spectral overlap of $98\pm0.4\%$ between the 18 sources, indicating that we have succeeded in fabricating many identical sources on a single chip. An example output spatial mode of the source is shown in the inset which has a $>98\%$ overlap with a single-mode fibre (Nufern 780-HP shown). }
\label{fig:sourceIllustration}
\end{center}
\end{figure*}

State-of-the-art experiments have investigated the quantum interference of light originating from three and four independent sources. Experiments of this scale, however, have required source imperfections to be compensated by narrow spectral filtering, long data collection times, and post-selected fidelity measures to show evidence of non-classical behavior~\cite{Yao2012a,Lu2007,Yao2012}. These current photon-source limitations have prohibited detailed studies generalising the HOM effect beyond two independent photons despite its wide-ranging utility in  a number of important technologies~\cite{Metcalf2013mqi,Spring2013bsp,Crespi2013a,Tillmann2014,Spagnolo2013a,Metcalf2014qtp}.

 Recently, integration of photon pair sources on-chip has been recognized as one of the most promising approaches to scaling due to their small size, direct compatibility with integrated photonic architectures, reduction in required pump power, and potentially exquisite control of the populated optical modes~\cite{Spring2013ocl,	Harris2014,Meany2014,Silverstone2013ocq}.  Unfortunately, fabrication imperfections or material limitations frequently spoil this dream.  Optical loss is a key parameter for any quantum light source and on-chip sources frequently suffer from large losses due to high scattering and outcoupling mode mismatch~\cite{Aboussouan2010hvt,Xiong2011gcp,Silverstone2013ocq}. In addition, the phase-matching conditions for the spontaneous scattering process are highly sensitive to optical dispersion. Even small non-uniformity in fabricating the waveguide structure can impart large changes in the joint spectrum of a generated photon pair. Further, poor dispersion control can introduce undesired residual frequency correlations between photon pairs which reduces heralded state purity, and thus the quality of interference. While recent experiments have managed to integrate two~\cite{Silverstone2013ocq} and four~\cite{Meany2014} individual photon pair sources on the same chip, they did not demonstrate the single photon purity necessary for performing heralded multi-photon quantum interference experiments.

Here we solve the problem of generating multiple independent \emph{pure-state} photons for the first time by means of a micro-fabricated waveguide array that contains 18 near-identical heralded single photon sources. We use our novel light source to map out, for the first time, the entire third-order correlation function arising from the interference of three independent single photons, showing genuine three-particle quantum interference.

\subsection*{Spontaneous Four-Wave Mixing in Silica}
\begin{figure*}[tb]
\centering
\includegraphics[width=0.9\textwidth]{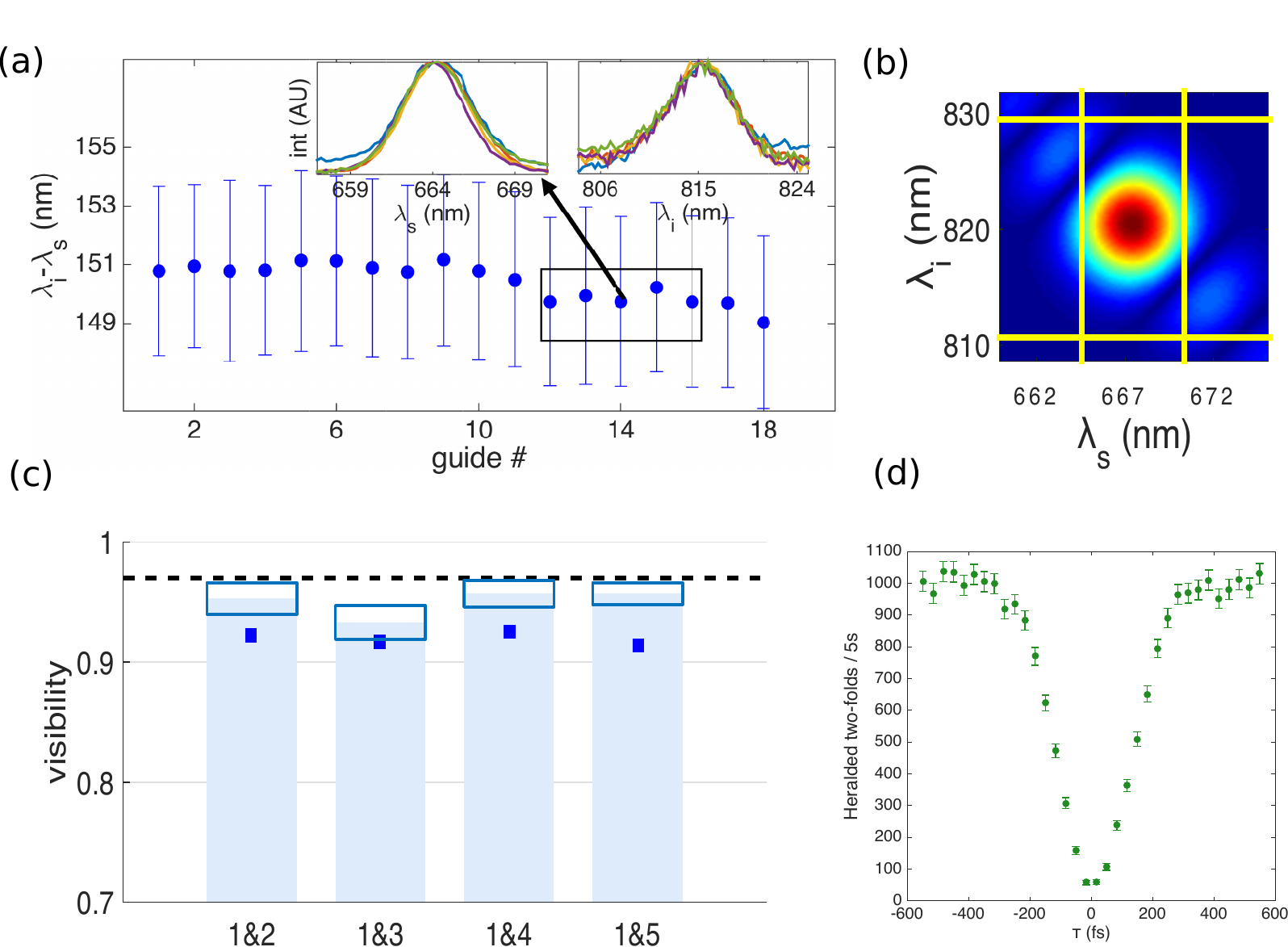}

\caption{{\bf Heralded HOM interference experiments} (a) The uniformity of UV-written SFWM sources is first quantified by measuring $\lambda_i-\lambda_s$ for $18$ different guides on the same chip. The full marginal spectra for each of the 18 idler photons are shown in figure~\ref{fig:sourceIllustration}. The errorbars represent the $1-\sigma$ widths of each marginal spectrum. The five sources shown in the box are selected for the two-source HOM-interference experiments. (b) The calculated Joint Spectral Amplitude (JSA) for our source. The yellow lines indicate the bandpass filters applied to each photon. This filtering removes residual spectral correlations between signal and idler and thus improves the heralded state purity from 87\% to 97\%, whilst still transmitting $>92\%$ of the generated single photons and without affecting the heralding efficiency. (c) Results of a series of HOM interference experiments using five waveguides, taking the heralded emission from two sources at a time.  Blue bars indicate background subtracted results, while squares show raw visibility results. The maximum expected visibility, $V^n_{max}=0.97$, is shown as a dashed line. This is due to the slight angular offset between the two herald photons providing a different bandpass from the shared spectral filters. This is a constraint of our bulk optics and is not intrinsic to the source; using dedicated filters for each individual source would restore $V^n_{max} > 0.99$.(d) Example two-source HOM interference data showing the reduction in heralded two-fold coincidences as the optical delay of one photon is scanned past the other.} \label{fig:2source}
\end{figure*}

Our source array consists of a series of straight waveguides fabricated by UV-laser writing of a silica-on-silicon photonic chip (figure \ref{fig:sourceIllustration}). The silica waveguides show low propagation loss and guide nearly-circular spatial modes (see inset to figure~\ref{fig:sourceIllustration}) that are nearly identical to those of a silica fibre. Each source is operated by injecting a pulse of horizontally polarized pump light. Spontaneous four-wave mixing (SFWM) is achieved through birefringent phase matching (see supplementary information) and generates the desired pairs of photons with vertical polarization. Appropriately matching the pump characteristics and phase-matching condition in the waveguides results in pairs of photons correlated only in their photon number, eliminating any excess frequency-time entanglement ~\cite{Smith2009ppg,Spring2013ocl}, thereby enabling heralded photons of high purity without the need for narrowband spectral or temporal filters. We use a nearly transform-limited pump pulse with a 4.5-nm bandwidth and 736-nm central wavelength to generate photons centered at 670\,nm and 817\,nm (see joint spectrum in figure~\ref{fig:2source}(b)). After the chip, the pump is removed with polarising optics and broadband spectral filters, while the photon pair is divided with a dichroic beam splitter and coupled into separate single-mode optical fibres (see supplementary figure~\ref{fig:expLayout})

Each source is individually characterized by monitoring the two output fibres with silicon avalanche photodiodes (APDs) while pump pulses of 1\,nJ pulse energy and $\sim160~\,{\rm fs}$ pulse duration are injected at a rate of 80~\,MHz. The heralding efficiency $\eta_{\mathrm{H}}$ is the probability that if a herald photon is detected, a photon in the heralded mode will be observed. We measure $\eta_{\mathrm{H}}=31\%$ (see supplementary information), for which the APDs are the dominant cause of inefficiency, and measure a coupling efficiency of $>80\,\%$ from the waveguide into single mode fibre making this source potentially useful for a range of loss-critical quantum applications~\cite{Varnava2008hgm,Datta2011qmi,Lucamarini2012die}. Although we observe some fluorescence noise in both modes of the source (see supplementary information), this can be almost entirely removed by time-gating our measurements to within $\sim1\,{\rm ns}$ of the pump. This effectively filters the time-uncorrelated fluorescence signal whilst not affecting the single photons that have a duration $\sim1000\times$ shorter than this gating window.  For the remaining work, the shorter-wavelength signal photon is used as the herald photon. We then measure an intensity autocorrelation $g^{(2)}_{\mathrm{H}} = 0.03$ of the heralded field, signifying good single-photon statistics, with the predominant error arising from multipair generation. At these conditions, heralded photons are detected at a rate of $\sim200\,{\rm kHz}$.

Whilst these excellent single-source characteristics are a pre-requisite for use in any quantum-enhanced  protocol, it is the construction of a large array of well-matched sources that has so far hindered the scaling of quantum photonic experiments. Critically, in the present device, we find the phase-matched wavelengths of the signal and idler photons for each of our sources are highly uniform, indicating a nearly constant birefringence across the entire silica-on-silicon wafer. The fabrication of the wafer by flame hydrolysis deposition is a well-developed technique widely used in classical photonics~\cite{Lepert2011}. This method achieves a uniformity of the substrate birefringence of $\delta(\Delta n)< 1\times10^{-6}$ which allows us to construct an unprecedented number of identical single photon sources on the same chip. 

As a exacting test that these waveguides are truly indistinguishable sources of pure single photons, a series of heralded two-source Hong-Ou-Mandel (HOM) interference experiments are performed pairwise between five different sources on the same chip, as shown in figure~\ref{fig:2source}. The output of each of four sources is interfered separately with a common reference source on a balanced fibre beam splitter to confirm near-identical, high purity emission among the entire set. An FPGA monitors the four-fold coincidences as an optical delay stage is scanned to record the HOM interference curves (see supplementary information). From the HOM curve we calculate the interference visibility as $V=(P^{(\infty)}-P^{(0)})/P^{(\infty)}$, where $P^{(\tau)}$ is the measured coincident probability at a time delay $\tau$. The experimentally observed interference visibilities closely match theoretical predictions in all cases while the similarity of raw and background-subtracted visibilities is consistent with minimal noise in heralded emission. We note that the five sources explored here form one representative subset of the 18 sources on the chip, highlighting the potential scalability of this architecture.

\begin{figure*}[ht]
\centering
\includegraphics[width=1\textwidth]{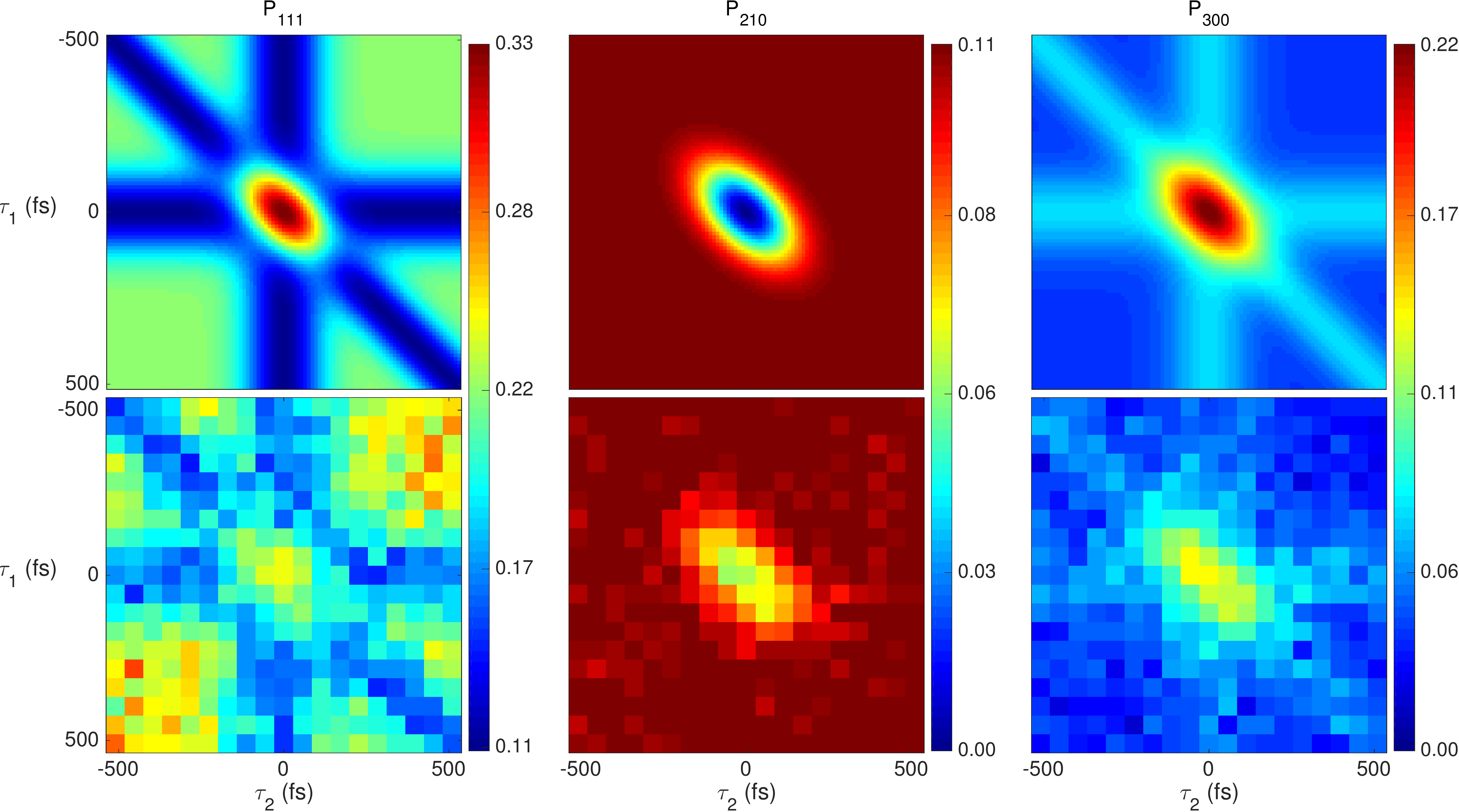}
\caption{{\bf  Experimental interference of three independent heralded photons in a tritter.} Top -  Theoretical output probabilities, $P_{111}$, $P_{210}$ and $P_{300}$ for ideal single Fock state inputs as a function of the delays of the impinging photons. Bottom -  Measured three photon interference data. The data constitute over $340,000$ six-photon events detected in $10$ days, a data rate of $0.41$ Hz, using commercial APDs.  Three heralded photons are coupled to a 3x3 fiber splitter with the resulting number statistics collected as two delay stages scan through the point of maximum indistinguishability (plot centers).  Pseudo-number resolution is obtained (for $N_{210}$ and $N_{300}$) by relying on additional fiber beam splitters and extra APDs.}
\label{fig:3source}
\end{figure*}

\subsection*{Interference of three heralded single photons}
This breakthrough in single photon source engineering; an array of identical, pure, heralded single photon sources  with sufficiently low-loss and high brightness to operate many of them simultaneously, allows us to go beyond previous studies of two-photon interference and explore the critically important regime of interference of multiple, independent single photons. This class of interference underpins all linear optical quantum communications, simulation and computing schemes, and obtaining high-fidelity operations based on interference is vital to understand the possibilities for scaling photonic systems to a useable capacity. Here, we show that our platform provides a route to large-scale photonics by performing an in-depth study of the entire interference landscape of three heralded photons for the first time. The quantum interference of three heralded single photons is studied by injecting the photons into a balanced three-port beam splitter (a fibre-based tritter). The action of this tritter is described by a unitary matrix that maps the input field operators, $\hat{a}^\dagger_i$ to the output field operators $\hat{b}^\dagger_i$ with $\hat{b}^\dagger_i=\sum_j\mathcal{U}_{i,j}\hat{a}^\dagger_j$, where the fibre tritter implements the unitary transformation:
\begin{equation}
  \label{eq:tritter}
  \mathcal{U}=\frac{1}{\sqrt{3}}\begin{pmatrix}
1 & 1 & 1 \\
1 & e^{i2\pi/3} & e^{i4\pi/3} \\
1 & e^{i4\pi/3} & e^{i2\pi/3}
 \end{pmatrix}.
\end{equation}
By varying the optical time-delay of the input states, we measure the full three-time third-order correlation function of the output photon statistics for each of the possible output combinations. Six-fold coincidence counts are recorded as a function of the two time delays using an array of silicon APDs and a custom FPGA-based counting unit. A second fibre-tritter is connected to one of the output modes to achieve pseudo photon-number resolution. The signal modes of each source pass through a broad spectral filter ($\Delta\lambda\sim10\,{\rm nm}$), the bandwidth of which is larger than the full width of the marginal spectra. This increases the purity of the heralded photons without significantly reducing the source brightness~\cite{Spring2013ocl}. The results of this three-photon generalisation of the Hong-Ou-Mandel experiment are shown in figure~\ref{fig:3source}. The average six-photon coincidence count rate is $0.41$ Hz~\footnote{This is the six-fold count rate through the fibre tritter including the second tritter used for resolving photon numbers at one of the output ports. The raw six-fold coincidence count rate into six single mode fibres is 2\,Hz.}, allowing us to collect enough data to resolve subtle features in the third-order correlations among modes after interference. 

\begin{figure}
  \centering
\includegraphics[width=0.9\columnwidth]{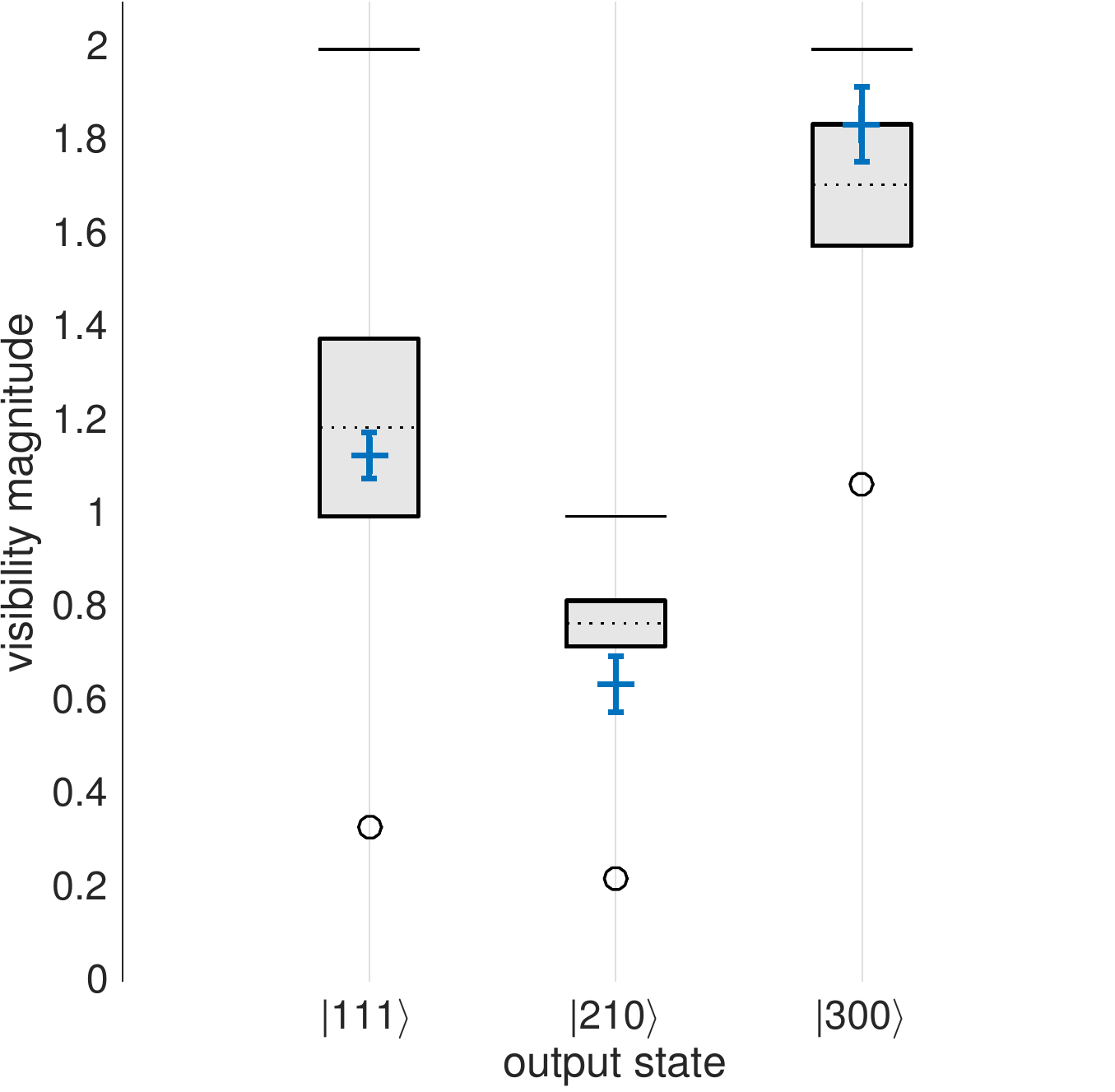}
  \caption{{\bf Measured and modeled interference visibilities}. For all results, $\tau_1=0$ and $\tau_2$ is scanned and the magnitude of the visibility is plotted. Our measured results are shown by the blue crosses. The shaded box represents the modeled visibility for three single photons, including the effects of higher-order pair emission and residual distinguishability. The errors on this model were estimated using a Monte-Carlo method taking into account the uncertainties in the model parameters.
Black solid lines indicate the maximum theoretical visibility for ideal Fock state inputs for which the full interference landscape is plotted in figure~\ref{fig:3source}. Black dashed lines indicate the maximum visibilities for classical coherent state inputs.}
  \label{fig:vis}
\end{figure}

The complicated structure of even these third-order correlation functions is indicative of the complexity in collective many-photon interference~\cite{Tichy2013}. To our knowledge this represents the first time such an entire third-order, three-time correlation surface has been mapped out experimentally. This allows us to observe directly the non-monotonic behaviour of the correlation functions as the distinguishability of the photons is changed, by adjusting the relative delays of the interfering photons, and can be seen in the plots of the $|111\rangle$ and $|300\rangle$ output terms. Such behaviour is realised only for quantum interference between more than 2 photons~\cite{Ra2013}. Further, this full correlation function clearly shows the distinction between two and three photon interference. In particular, we find the $|210\rangle$ output term is time-delay invariant if one photon always remains distinguishable. The complete absence of any two-photon interference effect is accompanied by a three-photon interference term which theoretically has unit visibility. We find this dip occurs when all photons arrive simultaneously (i.e. $\tau_1 = \tau_2 = 0$) and provides strong evidence for genuine three-photon interference. In fact, the suppression of this $|210\rangle$ term derives from the more general Fourier suppression law~\cite{Tichy2010} that has been suggested as a stringent test of the genuine quantum interference between many single photons~\cite{Tichy2013a}.

\subsection*{Discussion}
A full model was constructed to analyse the dependence of photon statistics on residual photon distinguishability, purity and higher-order SFWM pair emission (see supplementary information). The results from this model are shown in figure~\ref{fig:3source}. In order to quantify our experimentally observed three-photon visibilities, we ran a separate experiment setting $\tau_1=0$ and adjusted $\tau_2$ to introduce complete distinguishability. This one-dimensional scan allows collection of sufficient data to determine the three-photon interference contrast with high precision. The results are plotted in figure~\ref{fig:3source}. The expected visibilities for classical, phase-averaged, coherent state inputs are calculated and plotted as black dashed lines. The magnitudes of the measured visibilities clearly exceed these classical bounds. The experimentally measured effective squeezing parameter and source loss was used to model the effects of higher-order SFWM pair emission (see supplementary information). Further, the expected photon state purity as calculated from the theoretical joint spectral amplitude, together with the overlap of the measured marginal spectra, were used to include the effects of residual distinguishability and mixedness (see supplementary information). The results from this model are plotted as gray shaded boxes and agree very well with our measured results, confirming we have understood the primary sources of error. Importantly, we find that over $90\%$ of the visibility reduction is due to higher-order pair emission, which can be made arbitrarily small by reducing the pump power per source or using photon number resolving detectors on the herald arm. 

In conclusion, we have demonstrated a new platform for constructing large numbers of identical heralded single photon sources. The ability to operate these arrays of low-loss, high purity, on-chip sources will be increasingly critical as experiments scale to building and manipulating yet larger quantum states of light. This work has already allowed us to map out, for the first time, the three-photon generalisation of the Hong-Ou-Mandel interference landscape between independent photons. Our modeling indicates that the observed visibilities are a result of true tripartite interference and highlight the importance of minimizing loss, not only for increasing data rates, but in reducing the relative contribution of multi-pair emission to the observed click statistics~\cite{Bonneau2014}. 

Our results benefit both future source engineering and fundamental science efforts.  Such source arrays are an important component in engineering near-deterministic photon sources through active multiplexing~\cite{Migdall2002,Shapiro2007}. These would be useful in a wide range of quantum-enhanced technologies including metrology~\cite{Datta2011qmi}, communication~\cite{Gisin1991}, simulation~\cite{Aspuru-Guzik2012}, and computation.  However, even without multiplexing, an array of sources like that presented here is a valuable tool to probe the computational power of quantum systems~\cite{Aaronson2013, Lund2013, Bentivegna2015}.

\section*{Methods}
\subsection*{Waveguide fabrication}
\footnotesize{
The waveguide circuit used in this work was fabricated by the direct UV-writing technique utilizing silica slab waveguides deposited via flame hydrolysis deposition on a silicon substrate~\cite{Lepert2011}. The individual waveguides were written by focusing a continuous-wave UV laser (244 nm wavelength) onto the germanium doped silica photosensitive waveguide core and translating the laser beam transversely to the surface normal with computer-controlled 2D motion. The core layer is $4.8\,\mu{\rm m}$ thick with a step profile and the waveguides are laterally Gaussian in profile, formed as a result of UV-induced permanent refractive index change inside the photosensitive waveguide core layer. Single-mode operation was observed for pump, signal and idler photons with a typical mode field diameter of 5.5$\,\mu{\rm m}$ at the pump wavelength. The source waveguide propagation loss was measured to be 0.4\,dB/cm at the pump wavelength of $736\,$nm. The guides are $L=2.3$ cm long and are separated by either $50\,\mu{\rm m}$, $100\,\mu{\rm m}$ or $250\,\mu{\rm m}$ which confirms the ability to write many guides close to one another as well as being compatible with standard fibre-v-groove spacing for later integration. The three adjacent guides used for the three-photon experiments presented in this work were spaced by $50\,\mu{\rm m}$ and were coupled using a single aspheric lens. While dozens of guides fit on the chip, the results in this paper are drawn from a representative subset.

\subsection*{Spontaneous four-wave mixing in silica}
Phasematching SFWM in a birefringent waveguide is key to mitigating the main noise source in silica sources because the red frequency-detuned idler can be shifted far beyond silica's Raman gain peak\cite{Lin2007ppg,Smith2009ppg}. The consolidation process during silica layer fabrication results in a permanent stress between the planar layers due to the different coefficients of thermal expansion. This permanent stress manifests optically as a uniform birefringence across the entire wafer, naturally aligned with the chip axes. The magnitude of this birefringence can be controlled via the dopants incorporated into the glass. The birefringence of the chip used in this work was inferred to be $\Delta n = 1.25\times10^{-4}$. The  We quantify suppression of undesired higher-order photon emission via the heralded second-order coherence of the heralded mode, $g_H^{(2)}(0)$.  Perfect single photon sources achieve $g_H^{(2)}(0)=0$.  We measure $g_H^{(2)}(0)=0.08$ at the pump powers used for the six-photon experiments which is competitive with other heralded sources (see supplementary information). Fluorescence also contributes some noise but can be almost entirely removed with additional temporal and spectral filtering. These filters can be made very broad with respect to the single photon wavepacket and thus we can remove this noise without affecting the source loss (see supplementary information). Importantly, these high purity silica waveguides do not suffer from significant free carrier absorption which allows one to decouple source flux from source loss. In this way, it is possible to achieve high single photon production rates without inducing additional loss.

\subsection*{Spectral factorability}
 Undesired entanglement in the spectral degree of freedom leads to reduced purity in the heralded mode when broadband detectors are used.  Narrow spectral filters can be used to decrease detector bandwidth, and thus improve purity and interference contrast, but at the cost of reduced source flux.  Therefore, we wish to avoid narrow filters and instead emit directly into spectrally factorable signal and idler modes. The two photon component of the state emitted by our SFWM sources is
\begin{equation}
|\psi\rangle=\iint d\omega_sd\omega_if(\omega_s,\omega_i)|\omega_s\rangle|\omega_i\rangle
\label{eq:SFWMham}
\end{equation}
where the joint spectral amplitude $f(\omega_s,\omega_i)=\int d\omega'\alpha(\omega')\alpha(\omega_s+\omega_i-\omega')\phi(\omega_s,\omega_i)$ is a function of the pump envelope $\alpha(\omega)$ and phasematching function $\phi(\omega_s,\omega_i)=\text{sinc}(\Delta k L/2)\text{exp}(i\Delta k L/2)$ for a source length $L$ and $\Delta k=2k_p-k_s-k_i$ where $k_x=n_x\omega_x/c$ is the momentum of a photon with frequency $\omega_x$ in a dielectric of index $n_x$\cite{Garay-Palmett2007pps,Smith2009ppg}.  We wish to engineer spectrally factorable emission which requires $f(\omega_s,\omega_i)=g(\omega_s)h(\omega_i)$\cite{Grice2001efs}, where the exact forms of $g(\omega)$ and $h(\omega)$ are not important.  This factorable emission implies that the reduced state in the heralded mode, given detection of the herald, will be pure, which is a prerequisite for the high visibility interference demonstrated here. In silica, the group velocity of the pump lies between that of the signal and idler which allows near spectral factorability when $\Delta\lambda_p L = \gamma$ where $\gamma$ is a constant that depends on the dispersive properties of the dielectric.  For our sources with $L=2.3$ cm, we maximize spectral factorability with $\Delta\lambda_p=4.5$ nm.

\subsection*{Modeling interference visibilities}
We account for distinguishability and residual spectral mixedness among our three photons by using the multi-dimensional permanent of a tensor matrix\cite{Tichy2015spd}.  This method requires quantification of distinguishability and photon purity parameters. The photon state purity is estimated using the calculated Joint Spectral Amplitude as shown in figure~\ref{fig:2source}, taking account of the non-ideal placement of the spectral filters, from which we estimate a single source purity of 0.97. Using the background subtracted interference visibilities in figure \ref{fig:2source}, we estimate the remaining spectral distinguishability between the sources to be 0.98. Higher-order SFWM pair emission is modeled by iterating through all the ways in which a photon could be lost or multiple photons impinge on a single APD leading to an errant six-photon detection. Using the measured second-order correlation function and heralding efficiency (see supplementary information), we calculate $\lambda^2\sim0.025$, for our state, $\ket{\Psi}=\sqrt{(1-\lambda^2)}\sum_{n=0}^\infty\lambda^{n}\ket{nn}$. At each iteration, we determine the probability of the input leading to the specified output by calculating the permanent of a matrix that describes our lossy experiment.  This model offers the advantage of not requiring the full output state to be calculated, which includes many terms that cannot contribute to our data set.  While computation of permanents becomes intractable for larger matrices\cite{Aaronson2010ccl}, most current experiments operate at smaller scales where the model described here offers a relatively fast method to quantify the effects of both photon distinguishability and higher-order pair emission.}

\section*{Acknowlegements}
This work was supported by the UK Engineering and Physical Sciences Research Council (EPSRC EP/K034480/1, EP/M013243/1, and EP/J008052/1) and the European Office of Aerospace Research \& Development (AFOSR EOARD; FA8655-09-1-3020). W.S.K. is supported by an EC Marie Curie fellowship (PIEF-GA-2012-331859). B.J.S. is supported by the Oxford Martin Programme on Bio-Inspired Quantum Technologies. J.B.S. acknowledges support from the U.S. Air Force Institute of Technology. The views expressed in this article are those of the authors and do not reflect the official policy or position of the U.S. Air Force, Department of Defense, or the U.S. Government.
\section*{Author Contributions}
{\footnotesize J.B.S built and performed the experiments. J.B.S, B.J.M, P.C.H and W.S.K assisted with data-taking and analysed the data. B.J.M. performed the theoretical modeling. C.S., H.L.R, J.C.G and B.J.S performed initial feasibility studies with the source. P.L.M, H.L.R. and J.C.G designed, developed and fabricated the waveguide source.  J.B.S, B.J.S, P.G.R.S and I.A.W conceived the experiments. J.B.S and B.J.M wrote the manuscript with input from all authors. 
}

\pagebreak
\clearpage
\onecolumngrid

\makeatletter 

\renewcommand{\thefigure}{S\@arabic\c@figure} 
\renewcommand{\figurename}{Supplementary Figure}
\renewcommand{\thetable}{S\@arabic\c@table} 
\renewcommand{\tablename}{Supplementary Table}
\g@addto@macro\@floatboxreset\centering 

\def\tagform@#1{\maketag@@@{(S\ignorespaces#1\unskip\@@italiccorr)}} 

\makeatother

\setcounter{figure}{0}
\setcounter{equation}{0}

\section*{Supplementary Information: A chip-based array of near-identical, pure, heralded single photon sources}
\section{Experimental Setup}

\begin{figure*}[ht]
\begin{center}
\includegraphics[width=\textwidth]{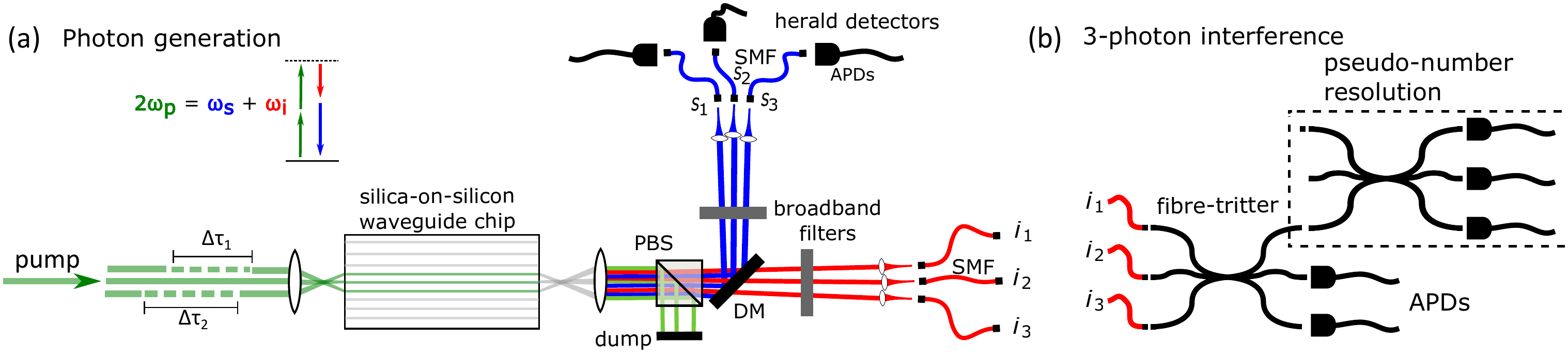}
\caption{\bf Experiments consist of (a) photon generation by heralded emission and (b) quantum interference.  (a) Sources are pumped by pulses from an $80$-MHz Ti-Sapphire laser oscillator centered at a wavelength $\lambda_p = 736$ nm.  A folded 4-f line with an adjustable slit (not shown) sets the pump bandwidth to $\Delta\lambda_p=4.5$ nm which maximizes spectral factorability.  In our three-source experiment, bulk beam splitters (not shown) split the pump with two beams transiting adjustable optical delay stages, $\Delta\tau_{1,2}$.  The three pump beams are efficiently coupled into separate on-chip waveguides through a single aspheric lens.  After the chip, a polarising beam splitter and broadband spectral filters remove the pump while a dichroic beam splitter separates signal and idler modes.  All modes are coupled to separate single-mode fibers.  (b)  Signal photons act as heralds and are coupled directly to APD detectors while idler photons interfere in a fiber beam splitter.  An FPGA (not shown) monitors coincident detection events on all detectors as the delays $\Delta\tau_{1,2}$ are scanned.  A second fiber three-port beam splitter is used to achieve pseudo-three-photon resolution with our binary APD detectors.  The two source experiments are similar except one pump path is blocked and a $2$-port fiber beam splitter is used for the interference experiments.}
\label{fig:expLayout}
\end{center}
\end{figure*}

\section{Noise sources and their mitigation}
\label{sec:noise}
We diagnose the principal source of noise from our sources and describe its mitigation with a combination of temporal and spectral filtering.  

\begin{figure}[htb]
\centering
\includegraphics[width = 0.6\textwidth]{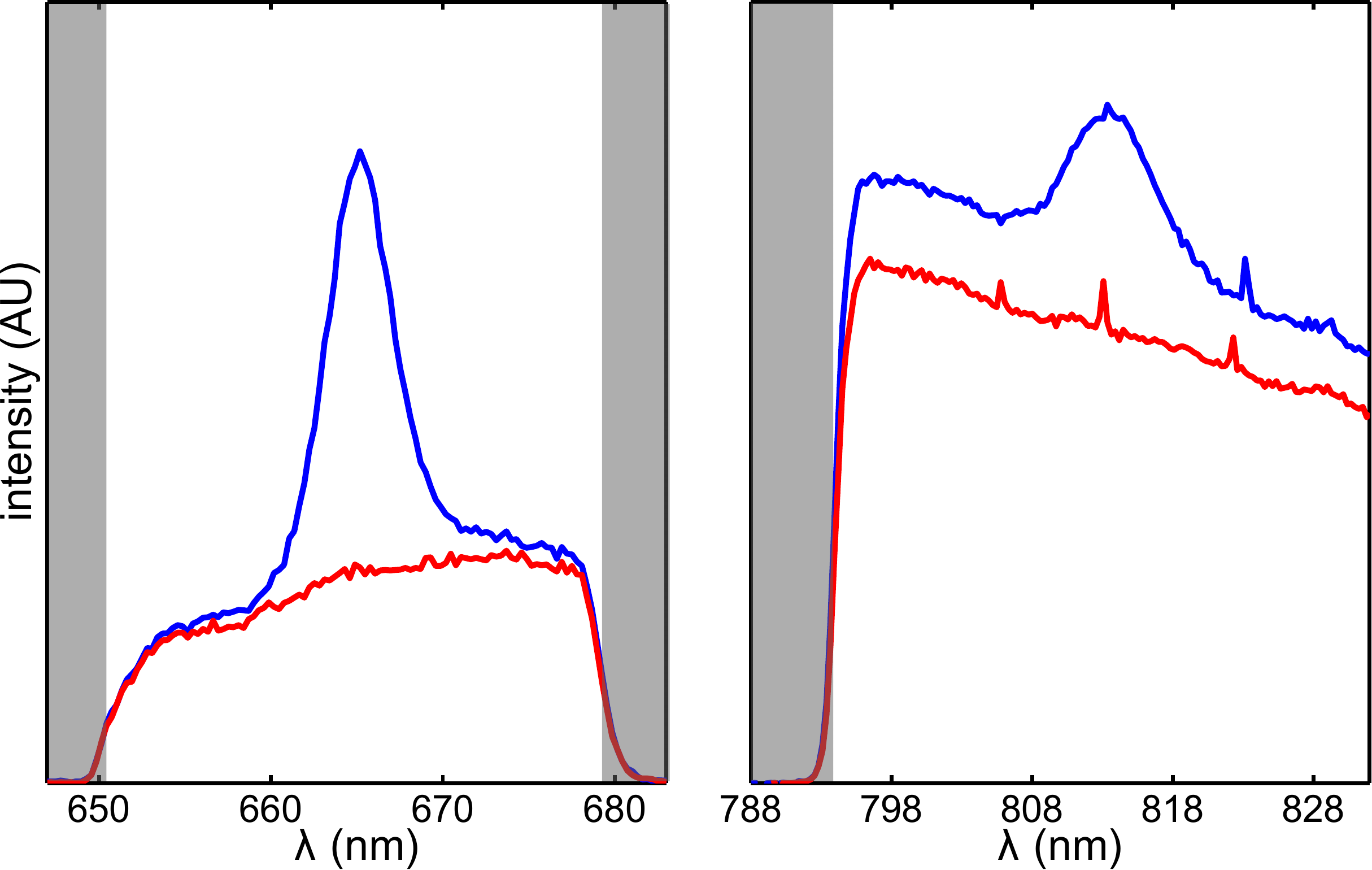}
\caption[Representative SFWM marginal spectra from UV-written waveguides]{A single photon spectrum shows the SFWM signal and idler at $\lambda_s=665$ nm and $\lambda_i=813$ nm respectively (blue) for a pump at $731$ nm (not shown).  There is significant background (red) on both the signal and idler modes.  Spectral filters attenuate frequencies in the gray shaded areas.  In the right plot, polarization dependent loss causes a systematic offset between the SFWM (blue) and background (red) in the idler mode.  In both the left and right plots, the background (red) extends significantly beyond the plotted regions and is limited here by the spectral filters (gray) and the spectrometer range.}
\label{c3fig:guide20marg}
\end{figure}

We apply a strong pulsed pump at $\lambda_p=736$ nm along the slow axis of our birefringent waveguide and observe the SFWM spectra shown in figure \ref{c3fig:guide20marg}.  By switching the pump polarisation to pump the fast axis (which is not phase-matched for SFWM), we can also record the background spectra. There is significant background evident for both the signal and idler modes shown in figure \ref{c3fig:guide20marg}.  This noise is not unique to this particular waveguide, but is representative of all the UV-written SFWM sources.  Fluorescence is the likely cause of this additional noise in our UV-written waveguides.

The noise overlaying the signal mode is potentially caused by the non-bridging oxygen-hole center (NBOHC) defect\cite{Adikan2008,Vaccaro2008lfn}.  This widely-studied silica defect is comprised of a dangling bond in the lattice $(\equiv\!\mathrm{Si}\!-\!\mathrm{O}\bullet)$ and can be caused by a number of physical mechanisms, including UV illumination \cite{Dianov1996ual}.  This defect exhibits a broad luminescence from $600-700$ nm with a peak at $655$ nm which lies close to the SFWM signal.  When our waveguides are pumped with a CW laser, the noise indicated by the red plot in the left of Figure \ref{c3fig:guide20marg} disappears (along with the SFWM).  Our hypothesis is that the high pump intensities when using an ultrashort pulsed pump allows for a multi-photon excitation of an NBOHC defect, which emits over a broad spectral region including that observed in Figure \ref{c3fig:guide20marg} overlapping the signal mode.

\begin{figure}[htb]
\centering
\includegraphics[width = 0.55\textwidth]{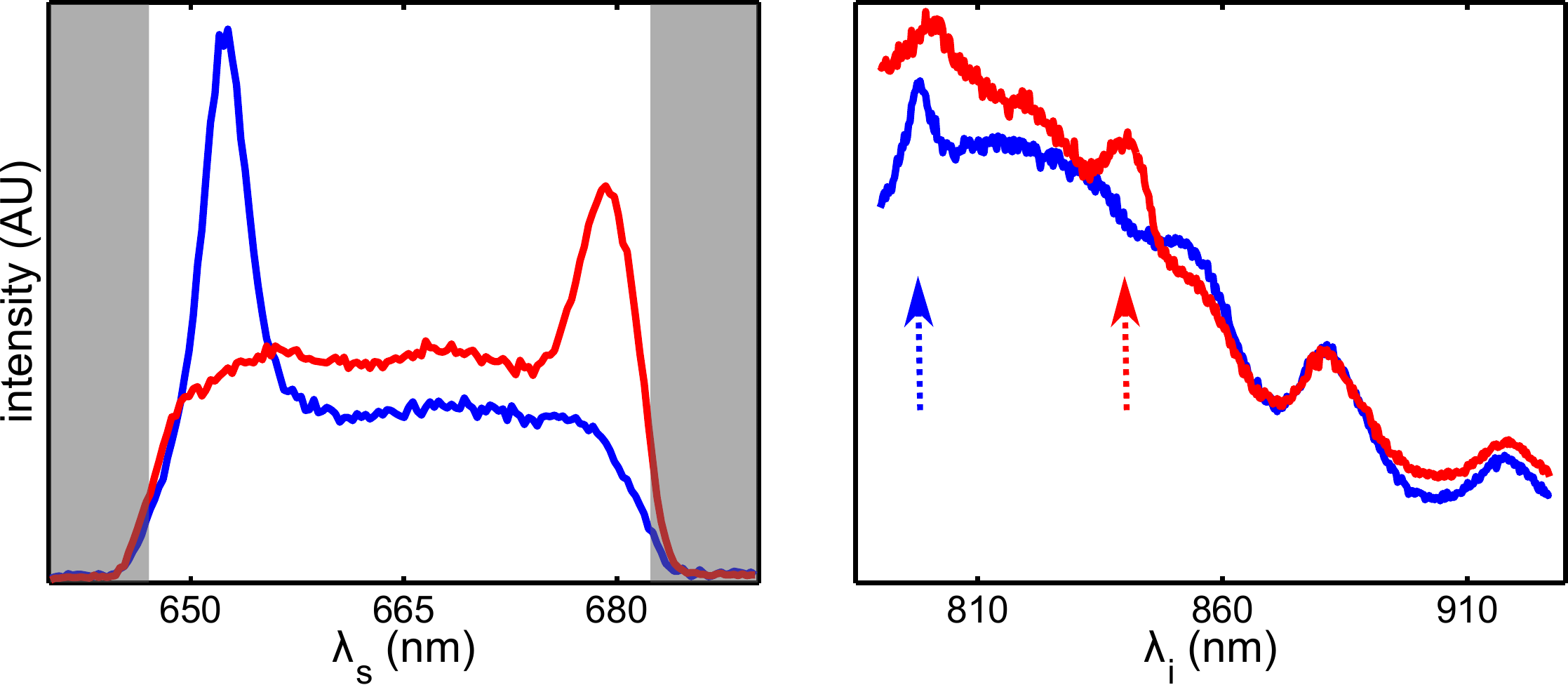}
\caption[Fluorescence noise in UV-written SFWM sources]{The spectra from a UV-written SFWM source is measured for two different pump wavelengths, $715$ nm (blue) and $750$ nm (red).  The signal mode is clearly visible on the left plot at the edges of the bandpass filter.  In the right plot, the idler is evident (arrows) on top of a broad background.  The shape of this background changes very little with the pump, which is consistent with fluorescence.}
\label{c3fig:margSpec2Pumps}
\end{figure}

The majority of the background surrounding the idler mode in Figure \ref{c3fig:guide20marg} is also fluorescence.  However, since it persists with CW pumping, it is not from the NBOHC defect described above.  While spontaneous Raman scattering has historically been a challenge for many SFWM pair sources, the $>40$ THz detuning of the idler from the pump implies that Raman scattering is minor.  This dominance of the noise by fluorescence, not Raman, is confirmed in Figure \ref{c3fig:margSpec2Pumps} where the noise spectrum is measured at two different pump wavelengths.  If the noise is due to fluorescence at a fixed frequency, then the noise spectrum should not change shape.  However, Raman emission exhibits a constant frequency detuning from the pump and thus should shift along with the pump.  The right side of Figure \ref{c3fig:margSpec2Pumps} demonstrates almost no change in the background at the two widely separated pump wavelengths, confirming the majority of the noise in these UV-written guides is due to fluorescence.  

Phasematching SFWM over a wide frequency range should, in principle, allow one to move the signal and idler frequencies to a range where this fluorescence background is minimized.  In practice, the currently available pump laser and detectors significantly constrain adjustability.  Our Ti-Sapphire pump laser cannot operate below $700$ nm while our silicon APDs monitoring the signal and idler can only operate in the range $400-1000$ nm.  For the multi-source experiments of interest here, we are further constrained to the most efficient region of the detectors from $500-850$ nm. We therefore consistently operate with wavelengths $\lambda_p\!\approx\! 730$ nm, $\lambda_s\!\approx\!665$ nm, and $\lambda_i\!\approx\! 815$ nm despite the presence of the fluorescence noise shown in Figure \ref{c3fig:margSpec2Pumps}.  Importantly, the proliferation of more efficient detectors at different spectral regions, particularly telecom wavelengths, will allow multi-source experiments in these spectral regions where fluorescence noise may not be a concern.

The background evident in Figure \ref{c3fig:guide20marg} could significantly deteriorate source performance.  Noise photons in the heralding mode reduce $\eta_H$ by introducing false herald events that are not correlated with any photons in the idler mode.  On the other hand, noise in the heralded mode causes undesired heralding of more than one photon that is reflected in an increased $g^{(2)}_H(0)$.  In both cases, noise reduces the photon number correlations between the signal and idler modes upon which our source performance relies.

\begin{figure}[htb]
\centering
\includegraphics[width = 0.55\textwidth]{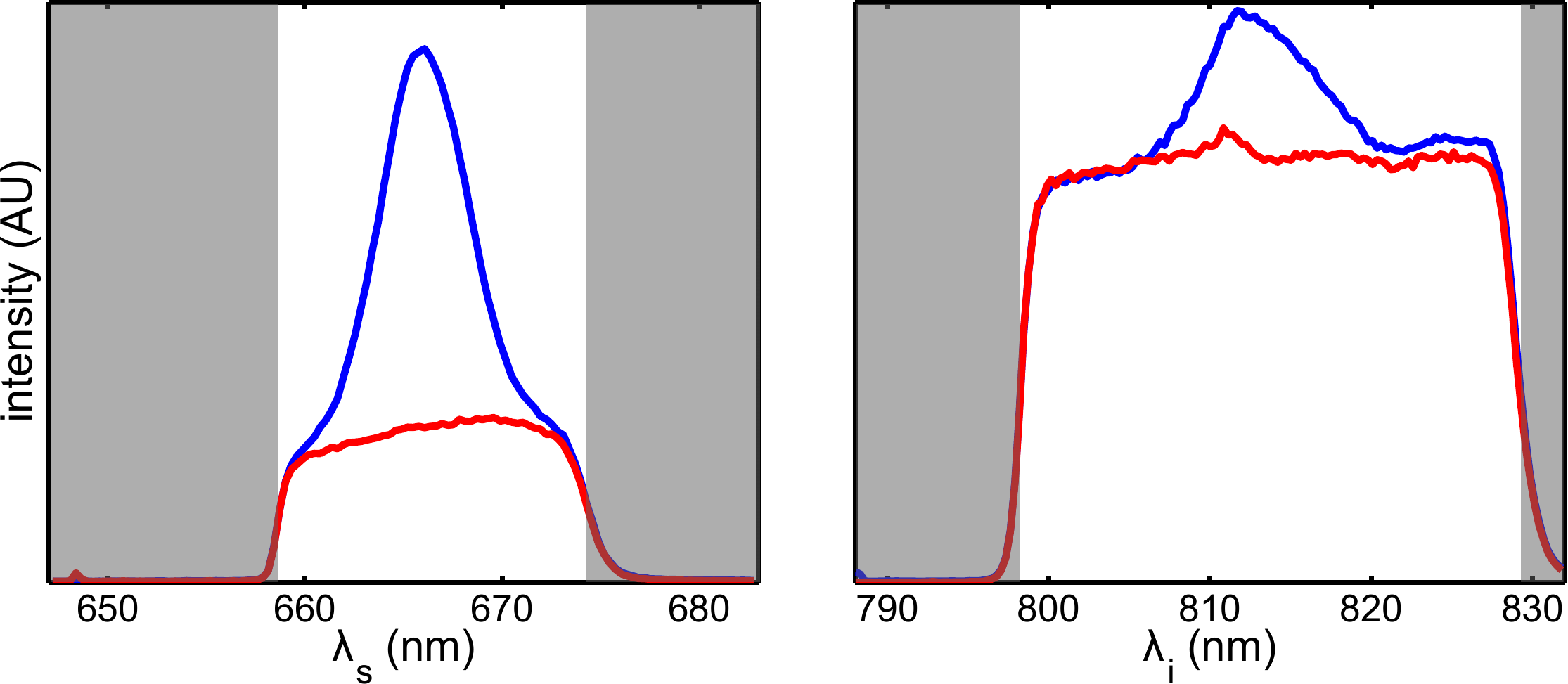}
\caption[Spectral filters angle-tuned to minimize noise]{The spectra of the SFWM signal and idler (blue) and corresponding background (red) are recorded before measuring the $\eta_H$ and $g^{(2)}_H(0)$ with APDs.  The spectral filters are angle-tuned to narrow the bandpass to minimize the background while introducing negligible loss on signal and idler.}
\label{c3fig:ORCmargSpecPreG2EtaH}
\end{figure}

Importantly, we are able to effectively filter this background without significantly attenuating the signal and idler modes, thus maximising source performance. First, we angle-tune our spectral filters to implement a bandpass filter about the signal and idler modes which is narrow with respect to the broad fluorescence features but broad enough to allow most ($>92\%$) of the signal and idler modes through. The results are shown in Figure \ref{c3fig:ORCmargSpecPreG2EtaH}, which is plotted on the same scale as the spectra using a larger bandpass in Figure \ref{c3fig:guide20marg}.  Second, we use a home-built FPGA-based coincidence counting program, with temporal resolution below its own clock period ($\sim\,400{\rm ps}$), to temporally filter the majority of the noise.  Fluorescence features in silica generally have decay times of microseconds which, for our $80$ MHz pulsed pump, implies a near temporally constant background.  Critically, the signal and idler wavepackets are of sub-picosecond duration which allows us to filter the temporally constant noise from the ultrashort pulsed SFWM by counting in coincidence with a pickoff from the pump laser. In this arrangement, a herald is a signal photon temporally coincident with a pump pulse.This removes the majority of the noise without affecting any of the SFWM.  

\begin{table}[htbp]
  \centering
  \caption[$\eta_H$ and $g^{(2)}_H(0)$ depends on pump power and degree of temporal filtering]{The heralding efficiency ($\eta_H$) is shown for a single SFWM source.  Each row corresponds to an average pump power, while the columns indicate the temporal window applied with the FPGA coincidence counting program. The two source experiments were run using a pump power of 80~\,mW whilst the three source experiments were run using a pump power of 135~\,mW. }
    \begin{tabular}{c|cccc}
\centering
               &$0.8$\,ns&$1.8$\,ns&$3.3$\,ns&$4.9$\,ns\\
        \hline
        $80$\,mW& $30.3\%$  & $28.3\%$  & $24.7\%$ & $22.7\%$\\
        $120$\,mW& $28.3\%$  & $28.6\%$  & $25.5\%$ & $23.8\%$\\
        $160$\,mW& $21.5\%$  & $28.1\%$ & $26.0\%$ & $24.3\%$\\      
    \end{tabular}%
      \label{tab:herEff}
\end{table}

\begin{table}
\centering
 \caption{The heralded second-order correlation $g^{(2)}_H(0)$ is shown for a single SFWM source. }
    \begin{tabular}{c|cccc}
\centering
		  &$0.8$\,ns& $1.8$\,ns&$3.3$\,ns&$4.9$\,ns\\
        \hline
         $80$\,mW& $0.032$  & $0.065$  &  $0.118$ & $0.154$\\
        $120$\,mW& $0.062$  & $0.099$  & $0.156$ & $0.196$\\
        $160$\,mW& $0.105$  & $0.144$ & $0.197$ & $0.248$\\      
    \end{tabular}%
  \label{tab:g2}
\end{table}

After this filtering, we analyze the number statistics of a single source to quantify $\eta_H$ and $g^{(2)}_H(0)$. Tables \ref{tab:herEff}-\ref{tab:g2} reveal that this combination of spectral and temporal filtering allows simultaneously high $\eta_H$ and low $g^{(2)}_H(0)$.  Table \ref{tab:g2} shows the expected behavior that increasing either the pump power or the width of the temporal filter increases $g^{(2)}_H(0)$ and thus increases the contribution of higher order photon emission.  The dependence of $\eta_H$ on these parameters in Table \ref{tab:herEff} is more complicated.  The highest value is obtained for the lowest pump power and tightest temporal filtering.  There isn't a monotonic decrease in $\eta_H$ for increasing powers and temporal windows since more noise photons in the heralded idler mode actually increase $\eta_H$.  However, the heralded state is not the desired single photon but rather a mixed state over the SFWM idler and noise.  In our experiments, we therefore implement tight temporal filtering and moderate pump powers to simultaneously attain high $\eta_H$ and low $g^{(2)}_H(0)$.  Minor changes to the optical alignment were made between the data taken in Tables \ref{tab:herEff}-\ref{tab:g2} and the main text.  While these tables still effectively quantify the effectiveness of spectral and temporal filtering, specific values for a given pump power and coincidence window are slightly different from those in the main text due to minor changes in optical alignment.

\section{Source homogeneity}

A goal of our experiment was to fabricate multiple identical SFWM sources on the same chip.  We measure the SFWM spectra from many guides on the same chip as one test to assess if this was achieved.  Source variation manifests as a change in the spectral position of the signal and idler.  Attempting to measure and compare just $\lambda_s$ or $\lambda_i$ between sources revealed errors dominated by small shifts in $\lambda_p$ between measurements.  We therefore measure $\lambda_i-\lambda_s$ which is insensitive to these small pump instabilities.  

\begin{figure}[htb]
\centering
\includegraphics[width = 0.7\textwidth]{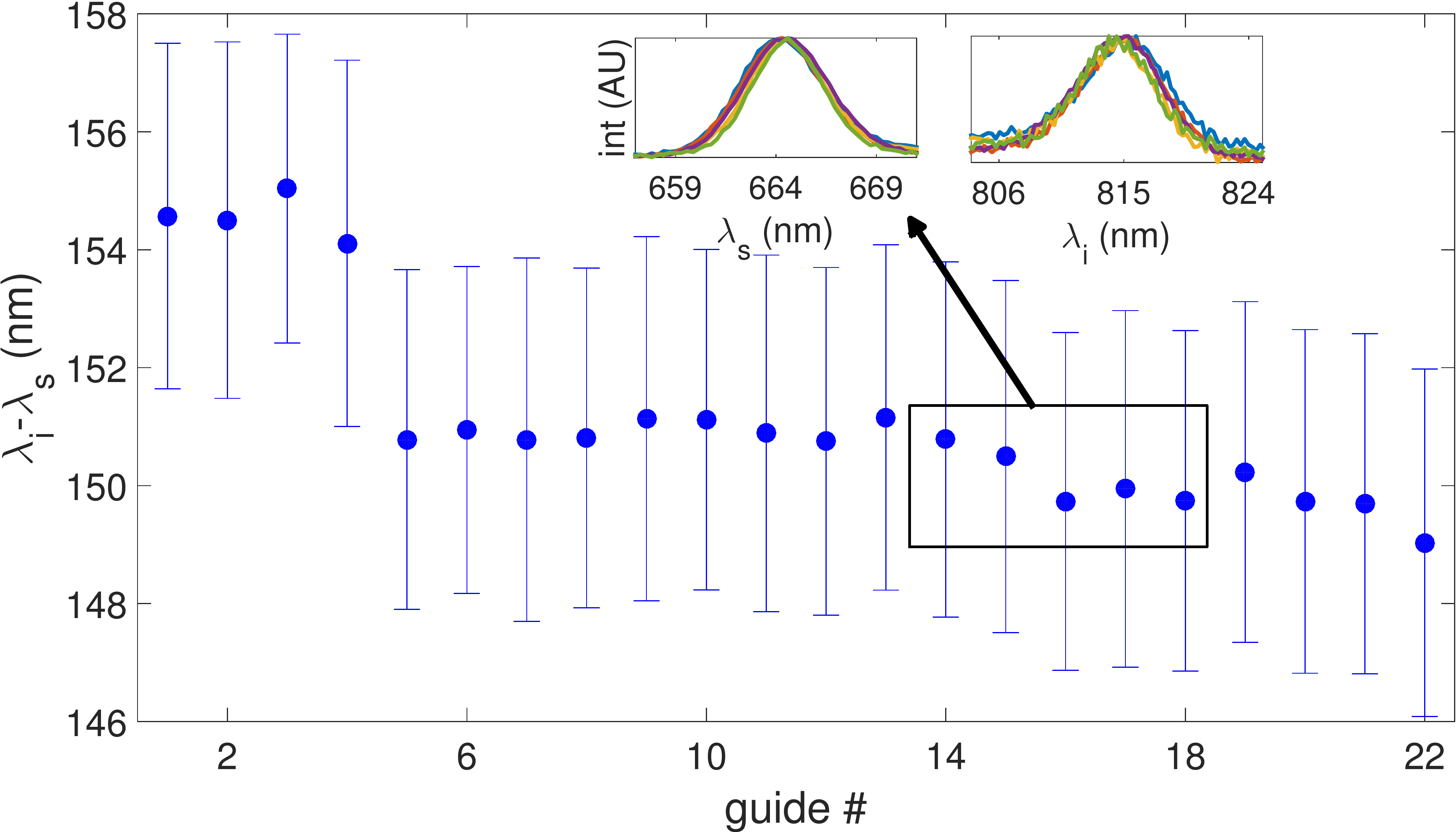}
\caption[Fabrication of matched UV-written SFWM sources]{The uniformity of UV-written SFWM sources is quantified by measuring $\lambda_i-\lambda_s$ for $22$ different guides on the same chip. The errorbars represent the $1-\sigma$ widths of each marginal spectrum. The phasematched wavelengths are determined by a Gaussian fit to the background subtracted data.  A representative set of background-subtracted data is shown in the insets for five of the sources.  All spectra are measured shortly after first exposure to the pulsed pump laser and are thus independent of the spectral shift effect discussed below. The main text discusses the results of guides 5-22 on this plot (18 waveguide sources total)}
\label{c3fig:uniformity40_3}
\end{figure}

Figure~\ref{c3fig:uniformity40_3} shows the results of this SFWM source homogeneity test.  The waveguides were UV-written in the order shown in Figure~\ref{c3fig:uniformity40_3}, with guide $1$ written first.  The majority of the waveguides are very closely matched. Performing a spectral overlap calculation between all possible pairs of sources yields a mean 91\% overlap with a standard deviation of 2.3\%. From the figure, we find that waveguides $1-4$ exhibited a much larger $\lambda_i-\lambda_s$ than the remaining guides. By excluding these waveguides we find the spectral overlap between the remaining 18 sources increases to $98\pm0.4\%$. The physical cause for this deviation amoung guides $1-4$ remains uncertain, but is rare and was not observed on other UV-written chips.  The remainder of the sources are well-matched with a slowly decreasing birefringence farther into the writing process.  There are at least two possible causes for this gradual drift.  Non-uniformity in the wafer from which this chip was diced could gradually change the stress, and thus the birefringence, across the chip.  Secondly, hydrogen is diffused into the chips to enhance their photosensitivity, this hydrogen gradually outgasses over the course of the writing processes causing the magnitude of the UV-induced index change to decrease over time.  These minor variations aside, the availability of many sources with nearly-identical spectra was more than sufficient for our experiment to proceed.  All results presented in the main text are from guides $5$-$9$.

\begin{figure}[htb]
\centering
\includegraphics[width = 0.6\textwidth]{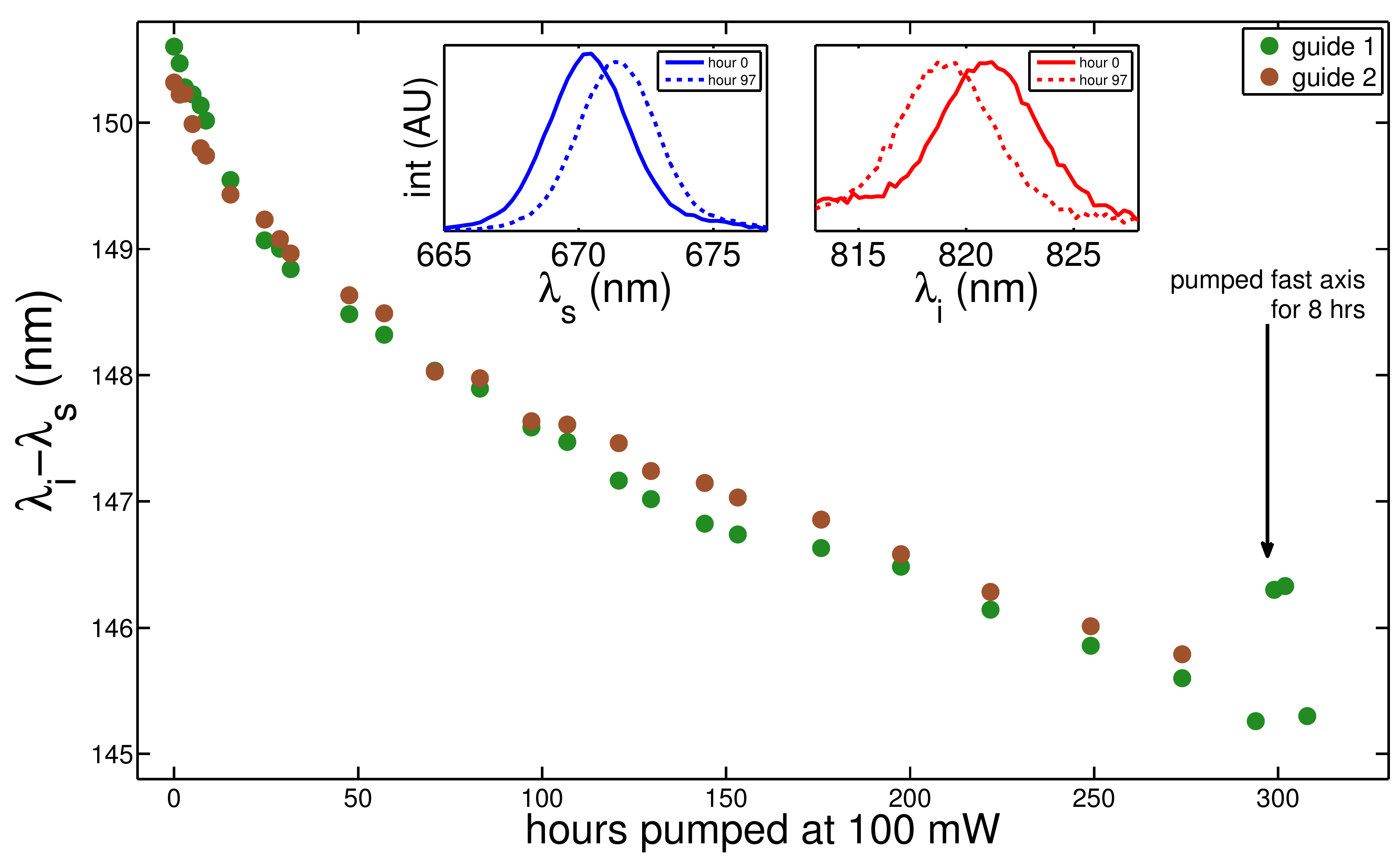}
\caption[Phasematched wavelengths from UV-written sources slowly shift]{The UV-written SFWM sources demonstrate a gradual shift in the signal and idler wavelength after being pumped for an extended period of time.  We quantify this shift by periodically measuring $\lambda_i-\lambda_s$ for two sources (brown/green) that are each pumped with the same average power ultrashort pulsed pump.  The insets show representative signal (blue) and idler (red) spectra at the start (solid) and finish (dotted) of the annealing run.  This spectral variation shows a polarization dependence; pumping the fast axis of the guide, where SFWM is not phase-matched, leads to an increase in $\lambda_i-\lambda_s$.}
\label{c3fig:anneal}
\end{figure}

Pumping our UV-written SFWM sources for extended periods of time causes a gradual decrease in the signal and idler detuning from the pump.  Figure~\ref{c3fig:anneal} shows this spectral shift for two waveguides, each of which are pumped with the same average power over the course of several days.  This spectral shift does not occur when the guides are pumped with a continuous wave source of equivalent power and center wavelength.  This indicates the shift is related to the high peak intensities present in the waveguide when an ultrashort pulsed pump is used.  We hypothesize that the proximity of our pump wavelength, $736$ nm, to a subharmonic of the absorption feature at 240-250nm targeted by the UV-writing process may play a role. It is possible that multi-photon absorption from the pulsed pump at $732$ nm may modify the photosensitive silica through the same mechanism as that used to fabricate the waveguides.

Before each interference experiment, we separately pump all involved sources until they spectrally match one another.  This procedure is used to correct for sources which are pumped different amounts of time during experimentation in order to replicate the inter-source homogeneity shown in Figure~\ref{c3fig:uniformity40_3} immediately after fabrication.  During subsequent periods of extended use, Figure~\ref{c3fig:anneal} demonstrates that the signal and idler spectra remain well-matched between multiple sources despite gradually shifting in time.  In other words, while the phasematched wavelengths slowly shift, they do so consistently from guide to guide and thus high-visibility interference experiments can run for days as sources remain spectrally matched throughout.  Importantly, we have noted that this spectral shift is not the limiting factor in stability when running sources for extended periods of time.  For example, spatial mode coupling and the optical path length matching degrade more quickly than the spectral match.  

The main motivation for pursuing UV-written SFWM sources was to achieve excellent inter-source homogeneity to allow high-visibility HOM interference.  The excellent initial source uniformity in Figure~\ref{c3fig:uniformity40_3} and the repeatable annealing demonstrated in Figure~\ref{c3fig:anneal} show that this goal is achievable

\section{Modeling interference visibilities}
\label{sec:modelling}

In this section we describe the numerical model used to simulate the expected click statistics from the fiber tritter considered in this work. There are three main sources of experimental error we account for in our model: photon distinguishability, photon state purity and higher-order SFWM pair emission. 

\subsection{Pure state distinguishability}
We first calculate the output state probabilities associated with the tritter when partially distinguishable photons are sent in. We use a method based on matrix permanents to calculate the transition probabilities associated with specific input and output states. Although this method has been used previously in the modeling of multi-photon interference~\cite{Metcalf2014qtp,Spring2013bsp}, in these cases the effects of partially distinguishable input photons were treated separately and added on to the model. In this work we instead implement a recently developed modification~\cite{Tichy2015spd} to this method which includes the effects of partial distinguishability from the beginning. 

Following the treatment presented in reference~\cite{Tichy2015spd}, we consider $n$ photons prepared in the input of a scattering setup characterized by a unitary $m\times m$ matrix $\mathcal{U}$. The initial distribution of particles is defined by the mode occupation list $\vec{r}=(r_1,\dots,r_m)$, where $r_j$ photons populate input mode $j$. Our aim is to calculate the probability that we will find the photons in a specific output state, $\vec{s}=(s_1,\dots,s_m)$. In a similar fashion to previous work involving the matrix permanent, unoccupied input and output ports have no effect on the calculated probabilities and so we need only work with the \emph{effective} scattering matrix, $M$, which is formed as a sub-matrix of $\mathcal{U}$ by taking rows and columns of the populated modes:
\begin{equation}
  \label{eq:M-matrix}
  M=\mathcal{U}_{\vec{d}(\vec{r}),\vec{d}(\vec{s})},
\end{equation}
where the mode assignment lists, $\vec{d}=(d_1,\dots,d_n)$ indicate the mode in which the $j$th photon resides. We now consider the effect of distinguishability between the input photons. We pool all possible distinguishing degrees of freedom of the photon in the $j$th input mode into an internal state, $\ket{\Phi_j}$ and assume all particles in the same input spatial mode are identical. Therefore, $\ket{\Phi_j}$ contains all the information that may potentially allow one to distinguish the photon in mode $j$ from another in mode $k$ other than the mode number itself - this includes all of frequency, time and polarisation. This mutual distinguishability information is encoded in the hermitian positive-definite $n\times n$ distinguishability matrix,
\begin{equation}
  \label{eq:S-matrix}
  \mathcal{S}_{j,k}=\braket{\Phi_{d_j(\vec{r})}}{\Phi_{d_k(\vec{r)}}}
\end{equation}
where $\mathcal{S}_{j,j} = 1$. With these definitions it can be shown~\cite{Tichy2015spd} that the probability of an output event $\vec{s}$ given an input state $\vec{r}$ described by a distinguishability matrix $\mathcal{S}$ is given by the multi-dimensional tensor permanent:
\begin{equation}
  \label{eq:multiPerm}
  \mathcal{P}_S(\vec{r},\vec{s}) = \mathcal{N}\times{\rm perm}(W) = \mathcal{N}\sum_{\sigma, \rho\in s_n}\prod_{j=1}^nW_{\sigma_j,\rho_j,j},
\end{equation}
where the $n^3$-dimensional three-tensor is defined as
\begin{equation}
  \label{eq:W-tensor}
  W_{k,l,j} = M_{k,j}M_{l,j}^*\mathcal{S}_{l,k},
\end{equation}
and $\sigma$, $\rho$ represent permutations of the output state, $\vec{s}$, and $A_{\sigma,\rho}^*$ denotes the complex conjugate of matrix $A$ with rows permuted according to $\sigma$ and columns permuted according to $\rho$. The normalisation factor, $\mathcal{N}=1/(\Pi_js_j!r_j!)$ is necessary to account for the possibility of multiply occupied input and output modes.

\subsection{Mixed state inputs}
This formalism can be straightforwardly extended to mixed input states by replacing the scalar products that appear in the matrix definition of the distinguishability matrix (given in equation~(\ref{eq:S-matrix})), by their ensemble average. We assume that the input photons are uncorrelated in any of their internal degrees of freedom (as is expected for three independently heralded photons) so that we can express each photon entering the $j$th occupied mode as being in the internal state $\ket{\Phi_{j,k}}$ with probability $p_{j,k}$,
\begin{equation}
  \label{eq:mixedState}
  \hat{\varrho}_j = \sum_{k=1}^Rp_{j,k}\proj{\Phi_{j,k}}{}{\Phi_{j,k}},
\end{equation}
with $R$ describing the maximal number of states in any pure-state decomposition across all input photons. The above expression for the event probability is then simply modified so that equation~(\ref{eq:multiPerm}) becomes:
\begin{equation}
  \label{eq:mixedPerm}
  \mathcal{P}_{[\hat{\vec{\varrho}}]}(\vec{r},\vec{s}) = \sum_{k_1,\dots, k_n=1}^R\left(\prod_{j=1}^np_{j,k_j}\right)\mathcal{P}_{\mathcal{S}[\vec{k}]}(\vec{r},\vec{s}),
\end{equation}
where $\mathcal{S}[\vec{k}]$ defines a particular $\mathcal{S}$-matrix given by:
\begin{equation}
  \label{eq:s-matrixVector}
  \mathcal{S}_{j,l}[\vec{k}]=\braket{\Phi_{j,k_j}}{\Phi_{l,k_l}}.
\end{equation}

\subsection{Higher-order pair emission}
Ideally, in our experiment, the input state remains fixed since we always want to inject three, single photon Fock states into each of the three input ports of the fibre tritter. Therefore we would set $\vec{r}=(1,1,1)$ ($\rightarrow \vec{d}=(1,2,3)$) and compute the probability for each of the three output terms we are interested in, namely, $\vec{s}=(1,1,1), \vec{s}=(2,1,0)$, and $\vec{s}=(3,0,0)$ ($\rightarrow \vec{d}=(1,2,3), \vec{d} = (1,1,2), \vec{d} = (1,1,1)$) using the expression given in equation~(\ref{eq:mixedPerm}).

However, we do not generate ideal single photon states since our sources are contaminated by higher-order photon pair terms. If we assume (for the purpose of this section) that our source is spectrally single mode, we can simply write down the two-mode squeezed state that is generated within our waveguided source:
\begin{equation}
  \label{eq:sfwm}
  \ket{\psi}_{\rm SFWM}=\sqrt{1-\vert\lambda\vert^2}\sum_{n=0}^\infty\lambda^n\ket{n_s,n_i}
\end{equation}
where $s$ and $i$ denote signal and idler modes and $\lambda$ is a generalised squeezing parameter determined by the pump power and strength of the $\chi^{(3)}$ non-linearity in silica. In our experiment, we herald on the signal mode in order to prepare approximate single photon Fock states in the idler mode. However, in practice our source, filters and accompanying optical setup have losses, and the heralding detector does not have unit efficiency and cannot resolve the number of photons in the signal mode. We let $\eta_s$ be the global loss on the heralding signal mode which accounts for all of these effects (including detector efficiency), and $\eta_i$ be the loss associated with the source on the idler mode - i.e. $\eta_i$ accounts for all losses from the source up until the fibre tritter. The effect of this loss (after tracing over the loss modes) is to transform the ideal state in equation~\eqref{eq:sfwm} into a binomial distribution over photon numbers~\cite{Bonneau2014}:
\begin{equation}
  \label{eq:SFWM:loss}
  \hat{\rho}_{\rm SFWM}=\left(1-\vert\lambda\vert^2\right)\left(\sum_{n=0}^\infty\vert\lambda\vert^{2n}\sum_{p,k,=0}^n\binom{n}{p}\eta_i^p(1-\eta_i)^{n-p}\binom{n}{k}\eta_s^k(1-\eta_s)^{n-k}\proj{p_i,k_s}{}{p_i,k_s}\right).
\end{equation}
Starting with this state we now use an APD to herald on the signal mode which reduces the state on the idler mode to:
\begin{equation}
  \label{eq:idelerMode}
  \hat{\rho}_i= \frac{{\rm tr}_s\left(\hat{\Pi}_{\rm APD}\hat{\rho}_{\rm SFWM}\right)}{{\rm tr}\left(\hat{\Pi}_{\rm APD}\hat{\rho}_{\rm SFWM}\right)},
\end{equation}
where the APD `click' POVM element is given by $\hat{\Pi}_{\rm APD}=\sum_{n=1}^\infty\proj{n}{}{n}$. We can write this out explicitly in terms of a classical mixture of input Fock states to our fibre tritter:
\begin{equation}
  \label{eq:inputstate}
  \hat{\rho}_i = \mathcal{A}\sum_{n=1}^\infty\vert\lambda\vert^{2n}\sum_{k=0}^n\mathcal{C}_{n,k}\proj{k_i}{}{k_i},
\end{equation}
where we have defined:
\begin{eqnarray}
  \label{eq:coeffsDef}
  \mathcal{A} &=& \frac{\left(1-\vert\lambda\vert^2\right)\left(1-\vert\lambda\vert^2\left(1-\eta_i\right)\right)}{\vert\lambda\vert^2\eta_i}, \\
\mathcal{C}_{n,k} &=& \sum_{p=1}^n\binom{n}{p}\eta_i^p(1-\eta_i)^{n-p}\binom{n}{k}\eta_s^k(1-\eta_s)^{n-k}.
\end{eqnarray}
The input to the fibre tritter consists of three such heralded states, $\rho_{in} = \rho_i^{(1)}\otimes\rho_i^{(2)}\otimes\rho_i^{(3)}$, so that the lowest order terms ($n=1$) in our input state contains terms that go as $\vert\lambda\vert^6$. In our model we keep all terms up to $\vert\lambda\vert^{10}$, which is sufficient to obtain an error of less than 1\% when $\lambda^2\sim0.01$ as is the case in our experiment. 

For example, expanding up to $n=3$ for each of the input terms and grouping photon numbers we have:
\begin{equation}
  \label{eq:expandRho}
  \hat{\rho}_i(n=3) = \mathcal{A}^3\left((\vert\lambda\vert^2\mathcal{C}_{1,0}+\vert\lambda\vert^4\mathcal{C}_{2,0}+\vert\lambda\vert^6\mathcal{C}_{3,0})\proj{0}{}{0}+(\vert\lambda\vert^2\mathcal{C}_{1,1}+\vert\lambda\vert^4\mathcal{C}_{2,1}+\vert\lambda\vert^6\mathcal{C}_{3,1})\proj{1}{}{1}+(\vert\lambda\vert^4\mathcal{C}_{2,2}+\vert\lambda\vert^6\mathcal{C}_{3,2})\proj{2}{}{2}+(\vert\lambda\vert^6\mathcal{C}_{3,3}\proj{3}{}{3})\right).
\end{equation}
When we compute the tensor product of this state to derive our three photon input we would be left with $4^3=64$ separate terms to compute. However, since we post-select on measuring three-photons after the tritter we can ignore all of those terms which contain less than a total of three photons at the input (since the probability for this event from equation~(\ref{eq:mixedPerm}) is trivially zero). We also ignore all of those terms of order $\vert\lambda\vert^{12}$ and above. This reduces the number of actual possible input states we must worry about to 22. Because we have assumed there are no correlations between the initial input modes, we simply perform the calculation for all possible 22 input states, and calculate all possible output state probabilities that could have led to a successful coincidence outcome using equation~(\ref{eq:mixedPerm}) to also account for the effects of modal distinguishability and mixedness. 

For example, one of the error terms from higher order emission leads to the input state $\ket{1,2,1}$ occasionally being input to the tritter when the central source emits a double pair. If we are calculating the probability for the output state to be $\ket{1,1,1}$, during the loop we then set the input state to $\vec{r}=(1,2,1)$ and use equation~(\ref{eq:mixedPerm}) to calculate the output probabilities for each of the terms which could lead to us detecting the $\ket{1,1,1}$ term: $\vec{s} = \left\{(2,1,1);(1,2,1);(1,1,2)\right\}$. We sum each of these contributions and then move on to the next possible input state, for example $\vec{r}=(2,1,0)$ and repeat. The final probability is then calculated by summing the contributions from each possible input state and weighting them by the appropriate coefficient given in equation~(\ref{eq:inputstate}).

\subsection{Experimental parameters for final model}
To use this model to make predictions of our experiment we need to make estimates for the various parameters that enter into the expressions derived in the previous section.

\begin{itemize}
\item {\bf Unitary matrix, $\mathcal{U}$:} We found that within our experimental precision (about 5\%), our fibre tritter implements the ideal operation and equally splits each input mode across all three output modes. We also assumed negligible loss within the fibre tritter since we were unable to accurately measure this independent of the connector losses on both input and output ports. Therefore we assumed the fibre tritter itself was well-described by a unitary matrix. All $3\times3$ unitary matrices can be fully characterised (up to unimportant input and output phases) using only knowledge of the splitting ratios of all input ports~\footnote{this is only true in the case of 2 or 3 mode devices - for unitary devices with four or more modes, there is an additional \emph{internal} phase freedom that means there are balanced 4-port devices, which do \emph{not} implement the discrete Fourier transform~\cite{Bernstein1974}}, which allows us to immediately write down our unitary transformation without having to characterise the internal phases:
  \begin{equation}
    \label{eq:tritterSI}
      \mathcal{U}=\frac{1}{\sqrt{3}}\begin{pmatrix}
1 & 1 & 1 \\
1 & e^{i2\pi/3} & e^{i4\pi/3} \\
1 & e^{i4\pi/3} & e^{i2\pi/3}
 \end{pmatrix}.
  \end{equation}
\item {\bf Photon purity matrix, $\mathcal{S}_{[k]}$:} Full quantum state tomography in every degree of freedom is difficult to implement experimentally for single photons, and in fact has never been fully demonstrated. Instead, we make a number of simplifying assumptions that allows us to make a good approximation to the description of the photon distinguishability / purity matrix $\mathcal{S}_{[k]}$. First, we assume each photon can be decomposed into a classical mixture across two orthogonal modes (i.e. setting $R=2$ in equation~(\ref{eq:mixedState})):
  \begin{equation}
    \label{eq:inputState_mixed}
    \hat{\rho}_{\rm in}=p\proj{\psi}{}{\psi}+(1-p)\ket{\psi}^{\perp\perp}\bra{\psi}
  \end{equation}
where $p$ is related to the purity of the photon through $\mathcal{P}={\rm tr}(\rho^2)=2p^2-2p+1$. Our first step is to estimate the purity of each photon, and so recover the coefficients required in equation~(\ref{eq:mixedPerm}). One particularly convenient method for measuring the purity is to make a measurement of the second-order auto-correlation of the idler mode, $g^{(2)}_{ii}(0)$~\cite{Mauerer2009hci}. A pure two-mode squeezed state will show thermal statistics ($g^{(2)}_{ii}(0)=2$) on its marginal modes. However, if there is any mixture (in any degree of freedom) this will show up as a decrease in the $g^{(2)}_{ii}(0)$ toward 1. 

Whilst other experiments have managed to use this method~\cite{Spring2013ocl,Mauerer2009hci}, we were not able to make accurate marginal measurements of $g^{(2)}_{ii}(0)$ because of the presence of uncorrelated noise on each of the marginal photons as explained in section~\ref{sec:noise}. With improved temporal filtering we anticipate being able to make this measurement on our source by counting in coincidence with the pump laser pulse.

Instead, for the purpose of this model, we use the theoretically expected purity. Since we use single spatial mode waveguides and perform polarisation filtering, we expect to have mixture only in the spectral degree of freedom. Using the known dispersion properties of silica, together with the measured birefringence of our waveguide and the shape and position of the spectral filters, we can compute the theoretical joint spectral amplitude for our source~\cite{Spring2014}. Taking the singular-value-decomposition of this function directly returns the expected source purity, which we calculate as $\mathcal{P}=0.97$. We further assume that this purity is identical for all three photons.

The entries of the $\mathcal{S}$-matrix depend only on the \emph{overlap} between the single photon states (see equation~(\ref{eq:S-matrix})) and so we let the first photon fully define the first basis mode (including spatial, spectral and polarisation modes) and parameterise the other two photons with respect to the first photon so that the six relevant photon wavefunctions become:

\begin{subequations}
  \label{eq:wavefunctions}
  \begin{align}
      \ket{\psi_1}&= (1,0)\quad\quad\quad\quad\quad\quad\,\,
  \ket{\psi_1}^\perp= (0,1)\\
  \ket{\psi_2} &= (\cos(\alpha), \sin(\alpha)) \quad\quad
  \ket{\psi_2}^\perp = (\sin(\alpha),-\cos(\alpha)) \\
\ket{\psi_3} &= (\cos(\beta),\sin(\beta)) \quad\quad
\ket{\psi_3}^\perp = (\sin(\beta),-\sin(\beta)).
  \end{align}
\end{subequations}
We can now estimate the parameters $\alpha$ and $\beta$ by looking at the visibility of the measured two-photon HOM dips presented in the main figure~\ref{fig:2source}. The visibility is given by $\mathcal{V}={\rm tr}(\rho_1\rho_2)$, and if we already know the purity of both photons (i.e. we know that ${\rm tr}(\rho_1^2)={\rm tr}(\rho_2^2)=\mathcal{P}=0.97$), then we can infer the value of $\alpha$ and $\beta$ from the value of visibility of the HOM-dips between photons 1 and 3 and photons 1 and 2 as is done in the main text. The relevant measurements are shown at the bottom of figure 4 as results `1\&4' and `1\&5' (since in this measurement, sources [`1',`2',`3',`4',`5'] correspond to waveguides [`20',`18',`19',`21',`22'] respectively. The three sources used in the main experiment are using guides 20,21 and 22). The background-subtracted visibility for both of these measurements is 0.957 leading to $\cos^2(\alpha)=\cos^2(\beta)=0.98$. 

We now have all the information needed to compute the $2^3=8$ necessary $\mathcal{S}$-matrices in equation~(\ref{eq:mixedPerm}). 

\item {\bf Signal mode loss, $\eta_s$:} Since we define the loss of the signal mode to include all the effects from photon production through to detection, a Klyshko-style efficiency measurement is all that is required~\cite{Klyshko1980}. Following the procedure outlined in ref~\cite{Eckstein}, we obtain an average efficiency on the signal arm of each source of $\eta_s=0.31\pm0.02$.

\item {\bf Idler mode loss, $\eta_i$:} This efficiency is only defined up until the point of the fibre tritter. Therefore we must remove the effects of detector efficiency from the Klyshko measurement of this channel efficiency. Using the manufacturers estimated efficiency of the detectors used in this arm we estimate that $\eta_i = 0.55\pm0.05$ (the increased uncertainty is due to our imperfect knowledge of the detector efficiency).
\item {\bf Squeezing parameter, $\lambda$:} A convenient property of the heralded second-order correlation function, $g^{(2)}_{si}(0)$, is that it is not affected by loss. We can make an easy measurement of this function using only a fibre beam splitter on one arm of the source. This is related to the squeezing parameter of the source through the simple relation~\cite{Spring2014}:
\begin{equation}
  \label{eq:16}
  \lambda^2 = g_{si}^{(2)}(0)\frac{\eta_s}{2(1-(1-\eta_s)^2)}.
\end{equation}

Using this relation and our knowledge of the herald arm efficiency, $\eta_s$ we determine that $\lambda^2=0.025\pm0.002$.
\end{itemize}

Using these estimated parameters we compute the expected three-photon probabilities for each of the output terms we are interested in at both the position of zero temporal delay, and when photon 1 is infinitely delayed by setting all entries of the $\mathcal{S}$-matrix containing the $\ket{\psi_1}$ terms to zero. We use these probabilities to compute the expected interference visibility from which we can directly compare to the experiment as we do in the main text figure~\ref{fig:3source}. An estimate of the error on this simulation is given by using a Monte-Carlo method to calculate 1000 instances of the probabilities, drawing the model parameters for each run from a Gaussian distribution centered at the measured values and with a standard deviation given by the errors on each model parameter.


\begin{thebibliography}{10}
\expandafter\ifx\csname url\endcsname\relax
  \def\url#1{\texttt{#1}}\fi
\expandafter\ifx\csname urlprefix\endcsname\relax\def\urlprefix{URL }\fi
\providecommand{\bibinfo}[2]{#2}
\providecommand{\eprint}[2][]{\url{#2}}

\bibitem{URen2003}
\bibinfo{author}{U'Ren, A.~B.}, \bibinfo{author}{Mukamel, E.},
  \bibinfo{author}{Banaszek, K.} \& \bibinfo{author}{Walmsley, I.~a.}
\newblock \bibinfo{title}{{Managing photons for quantum information
  processing.}}
\newblock \emph{\bibinfo{journal}{Philosophical transactions. Series A,
  Mathematical, physical, and engineering sciences}}
  \textbf{\bibinfo{volume}{361}}, \bibinfo{pages}{1493--1506}
  (\bibinfo{year}{2003}).

\bibitem{Zukowski1995}
\bibinfo{author}{Zukowski, M.}, \bibinfo{author}{Zeilinger, A.} \&
  \bibinfo{author}{Weinfurter, H.}
\newblock \bibinfo{title}{{Entangling Photons Radiated by Independent Pulsed
  Sources}}.
\newblock \emph{\bibinfo{journal}{Annals of the New York Academy of Sciences}}
  \textbf{\bibinfo{volume}{755}}, \bibinfo{pages}{91--102}
  (\bibinfo{year}{1995}).

\bibitem{Walmsley2005a}
\bibinfo{author}{Walmsley, I.~a.} \& \bibinfo{author}{Raymer, M.~G.}
\newblock \bibinfo{title}{{Applied physics. Toward quantum-information
  processing with photons.}}
\newblock \emph{\bibinfo{journal}{Science (New York, N.Y.)}}
  \textbf{\bibinfo{volume}{307}}, \bibinfo{pages}{1733--4}
  (\bibinfo{year}{2005}).

\bibitem{Knill2001}
\bibinfo{author}{Knill, E.}, \bibinfo{author}{Laflamme, R.} \&
  \bibinfo{author}{Milburn, G.}
\newblock \bibinfo{title}{{A scheme for efficient quantum computation with
  linear optics}}.
\newblock \emph{\bibinfo{journal}{Nature}} \textbf{\bibinfo{volume}{409}},
  \bibinfo{pages}{46--52} (\bibinfo{year}{2001}).

\bibitem{Intervals1987}
\bibinfo{author}{Hong, C.}, \bibinfo{author}{Ou, Z.} \&
  \bibinfo{author}{Mandel, L.}
\newblock \bibinfo{title}{{Measurement of Subpicosecond Time Intervals between
  Two Photons by Interference}}.
\newblock \emph{\bibinfo{journal}{Physical Review Letters}}
  \textbf{\bibinfo{volume}{59}}, \bibinfo{pages}{2044--2046}
  (\bibinfo{year}{1987}).

\bibitem{Calkins2013}
\bibinfo{author}{Calkins, B.} \emph{et~al.}
\newblock \bibinfo{title}{{High quantum-efficiency photon-number-resolving
  detector for photonic on-chip information processing}}.
\newblock \emph{\bibinfo{journal}{Optics Express}}
  \textbf{\bibinfo{volume}{21}}, \bibinfo{pages}{22657--22670}
  (\bibinfo{year}{2013}).

\bibitem{Carolan2015}
\bibinfo{author}{Carolan, J.} \emph{et~al.}
\newblock \bibinfo{title}{{Universal linear optics}}.
\newblock \emph{\bibinfo{journal}{Science}} \textbf{\bibinfo{volume}{349}},
  \bibinfo{pages}{11--16} (\bibinfo{year}{2015}).

\bibitem{Metcalf2014qtp}
\bibinfo{author}{Metcalf, B.~J.} \emph{et~al.}
\newblock \bibinfo{title}{{Quantum teleportation on a photonic chip}}.
\newblock \emph{\bibinfo{journal}{Nature Photonics}}
  \textbf{\bibinfo{volume}{8}}, \bibinfo{pages}{770--774}
  (\bibinfo{year}{2014}).

\bibitem{Heshami2015}
\bibinfo{author}{Heshami, K.} \emph{et~al.}
\newblock \bibinfo{title}{{Quantum memories: emerging applications and recent
  advances}}.
\newblock \emph{\bibinfo{journal}{arXiv}} \textbf{\bibinfo{volume}{00}},
  \bibinfo{pages}{1--32} (\bibinfo{year}{2015}).

\bibitem{Eisaman2011}
\bibinfo{author}{Eisaman, M.~D.}, \bibinfo{author}{Fan, J.},
  \bibinfo{author}{Migdall, a.} \& \bibinfo{author}{Polyakov, S.~V.}
\newblock \bibinfo{title}{{Invited review article: Single-photon sources and
  detectors.}}
\newblock \emph{\bibinfo{journal}{The Review of scientific instruments}}
  \textbf{\bibinfo{volume}{82}}, \bibinfo{pages}{071101}
  (\bibinfo{year}{2011}).

\bibitem{Migdall2002}
\bibinfo{author}{Migdall, A.}, \bibinfo{author}{Branning, D.} \&
  \bibinfo{author}{Castelletto, S.}
\newblock \bibinfo{title}{{Tailoring single-photon and multiphoton
  probabilities of a single-photon on-demand source}}.
\newblock \emph{\bibinfo{journal}{Physical Review A}}
  \textbf{\bibinfo{volume}{66}}, \bibinfo{pages}{053805}
  (\bibinfo{year}{2002}).

\bibitem{Yao2012a}
\bibinfo{author}{Yao, X.-C.} \emph{et~al.}
\newblock \bibinfo{title}{{Experimental demonstration of topological error
  correction}}.
\newblock \emph{\bibinfo{journal}{Nature}} \textbf{\bibinfo{volume}{482}},
  \bibinfo{pages}{489--494} (\bibinfo{year}{2012}).

\bibitem{Lu2007}
\bibinfo{author}{Lu, C.-Y.} \emph{et~al.}
\newblock \bibinfo{title}{{Experimental entanglement of six photons in graph
  states}}.
\newblock \emph{\bibinfo{journal}{Nature Physics}}
  \textbf{\bibinfo{volume}{3}}, \bibinfo{pages}{91--95} (\bibinfo{year}{2007}).

\bibitem{Yao2012}
\bibinfo{author}{Yao, X.-C.} \emph{et~al.}
\newblock \bibinfo{title}{{Observation of eight-photon entanglement}}.
\newblock \emph{\bibinfo{journal}{Nature Photonics}}
  \textbf{\bibinfo{volume}{6}}, \bibinfo{pages}{225--228}
  (\bibinfo{year}{2012}).

\bibitem{Metcalf2013mqi}
\bibinfo{author}{Metcalf, B.~J.} \emph{et~al.}
\newblock \bibinfo{title}{{Multiphoton quantum interference in a multiport
  integrated photonic device.}}
\newblock \emph{\bibinfo{journal}{Nature Communications}}
  \textbf{\bibinfo{volume}{4}}, \bibinfo{pages}{1356} (\bibinfo{year}{2013}).

\bibitem{Spring2013bsp}
\bibinfo{author}{Spring, J.~B.} \emph{et~al.}
\newblock \bibinfo{title}{{Boson Sampling on a Photonic Chip}}.
\newblock \emph{\bibinfo{journal}{Science (New York, N.Y.)}}
  \textbf{\bibinfo{volume}{339}}, \bibinfo{pages}{798--801}
  (\bibinfo{year}{2013}).

\bibitem{Crespi2013a}
\bibinfo{author}{Crespi, A.} \emph{et~al.}
\newblock \bibinfo{title}{{Integrated multimode interferometers with arbitrary
  designs for photonic boson sampling}}.
\newblock \emph{\bibinfo{journal}{Nature Photonics}}
  \textbf{\bibinfo{volume}{7}}, \bibinfo{pages}{545--549}
  (\bibinfo{year}{2013}).

\bibitem{Tillmann2014}
\bibinfo{author}{Tillmann, M.} \emph{et~al.}
\newblock \bibinfo{title}{{BosonSampling with Controllable
  Distinguishability}}.
\newblock \emph{\bibinfo{journal}{arXiv preprint arXiv:1403.3433}}
  (\bibinfo{year}{2014}).

\bibitem{Spagnolo2013a}
\bibinfo{author}{Spagnolo, N.} \emph{et~al.}
\newblock \bibinfo{title}{{Three-photon bosonic coalescence in an integrated
  tritter.}}
\newblock \emph{\bibinfo{journal}{Nature communications}}
  \textbf{\bibinfo{volume}{4}}, \bibinfo{pages}{1606} (\bibinfo{year}{2013}).

\bibitem{Spring2013ocl}
\bibinfo{author}{Spring, J.~B.} \emph{et~al.}
\newblock \bibinfo{title}{{On-chip low loss heralded source of pure single
  photons}}.
\newblock \emph{\bibinfo{journal}{Optics Express}}
  \textbf{\bibinfo{volume}{21}}, \bibinfo{pages}{5932--5935}
  (\bibinfo{year}{2013}).

\bibitem{Harris2014}
\bibinfo{author}{Harris, N.~C.} \emph{et~al.}
\newblock \bibinfo{title}{{Integrated source of spectrally filtered correlated
  photons for large-scale quantum photonic systems}}.
\newblock \emph{\bibinfo{journal}{Physical Review X}}
  \textbf{\bibinfo{volume}{4}}, \bibinfo{pages}{1--10} (\bibinfo{year}{2014}).

\bibitem{Meany2014}
\bibinfo{author}{Meany, T.} \emph{et~al.}
\newblock \bibinfo{title}{{Hybrid photonic circuit for multiplexed heralded
  single photons}}.
\newblock \emph{\bibinfo{journal}{Laser and Photonics Reviews}}
  \textbf{\bibinfo{volume}{8}}, \bibinfo{pages}{42--46} (\bibinfo{year}{2014}).

\bibitem{Silverstone2013ocq}
\bibinfo{author}{Silverstone, J.~W.} \emph{et~al.}
\newblock \bibinfo{title}{{On-chip quantum interference between silicon
  photon-pair sources}}.
\newblock \emph{\bibinfo{journal}{Nature Photonics}} \bibinfo{pages}{2--6}
  (\bibinfo{year}{2013}).

\bibitem{Aboussouan2010hvt}
\bibinfo{author}{Aboussouan, P.}, \bibinfo{author}{Alibart, O.},
  \bibinfo{author}{Ostrowsky, D.~B.}, \bibinfo{author}{Baldi, P.} \&
  \bibinfo{author}{Tanzilli, S.}
\newblock \bibinfo{title}{{High-visibility two-photon interference at a telecom
  wavelength using picosecond-regime separated sources}}.
\newblock \emph{\bibinfo{journal}{Physical Review A}}
  \textbf{\bibinfo{volume}{81}} (\bibinfo{year}{2010}).

\bibitem{Xiong2011gcp}
\bibinfo{author}{Xiong, C.} \emph{et~al.}
\newblock \bibinfo{title}{{Generation of correlated photon pairs in a
  chalcogenide As[sub 2]S[sub 3] waveguide}}.
\newblock \emph{\bibinfo{journal}{Applied Physics Letters}}
  \textbf{\bibinfo{volume}{98}}, \bibinfo{pages}{51101} (\bibinfo{year}{2011}).

\bibitem{Smith2009ppg}
\bibinfo{author}{Smith, B.~J.}, \bibinfo{author}{Mahou, P.},
  \bibinfo{author}{Cohen, O.}, \bibinfo{author}{Lundeen, J.~S.} \&
  \bibinfo{author}{Walmsley, I.~A.}
\newblock \bibinfo{title}{{Photon pair generation in birefringent optical
  fibers}}.
\newblock \emph{\bibinfo{journal}{Optics Express}}
  \textbf{\bibinfo{volume}{17}}, \bibinfo{pages}{23589} (\bibinfo{year}{2009}).

\bibitem{Varnava2008hgm}
\bibinfo{author}{Varnava, M.}, \bibinfo{author}{Browne, D.} \&
  \bibinfo{author}{Rudolph, T.}
\newblock \bibinfo{title}{{How Good Must Single Photon Sources and Detectors Be
  for Efficient Linear Optical Quantum Computation?}}
\newblock \emph{\bibinfo{journal}{Physical Review Letters}}
  \textbf{\bibinfo{volume}{100}}, \bibinfo{pages}{60502}
  (\bibinfo{year}{2008}).

\bibitem{Datta2011qmi}
\bibinfo{author}{Datta, A.} \emph{et~al.}
\newblock \bibinfo{title}{{Quantum metrology with imperfect states and
  detectors}}.
\newblock \emph{\bibinfo{journal}{Physical Review A}}
  \textbf{\bibinfo{volume}{83}}, \bibinfo{pages}{63836} (\bibinfo{year}{2011}).

\bibitem{Lucamarini2012die}
\bibinfo{author}{Lucamarini, M.}, \bibinfo{author}{Vallone, G.},
  \bibinfo{author}{Gianani, I.}, \bibinfo{author}{Mataloni, P.} \&
  \bibinfo{author}{{Di Giuseppe}, G.}
\newblock \bibinfo{title}{{Device-independent entanglement-based Bennett 1992
  protocol}}.
\newblock \emph{\bibinfo{journal}{Physical Review A}}
  \textbf{\bibinfo{volume}{86}}, \bibinfo{pages}{32325} (\bibinfo{year}{2012}).

\bibitem{Lepert2011}
\bibinfo{author}{Lepert, G.} \emph{et~al.}
\newblock \bibinfo{title}{{Demonstration of UV-written waveguides, Bragg
  gratings and cavities at 780 nm, and an original experimental measurement of
  group delay}}.
\newblock \emph{\bibinfo{journal}{Optics Express}}
  \textbf{\bibinfo{volume}{19}}, \bibinfo{pages}{24933--24943}
  (\bibinfo{year}{2011}).

\bibitem{Note1}
\bibinfo{note}{This is the six-fold count rate through the fibre tritter
  including the second tritter used for resolving photon numbers at one of the
  output ports. The raw six-fold coincidence count rate into six single mode
  fibres is 2\protect \tmspace +\thinmuskip {.1667em}Hz.}

\bibitem{Tichy2013}
\bibinfo{author}{Tichy, M.}
\newblock \bibinfo{title}{{Interference of Identical Particles from
  Entanglement to Boson-Sampling}}.
\newblock \emph{\bibinfo{journal}{J. Phys. B: At. Mol. Opt. Phys}}
  \textbf{\bibinfo{volume}{47}}, \bibinfo{pages}{103001}
  (\bibinfo{year}{2014}).

\bibitem{Ra2013}
\bibinfo{author}{Ra, Y.~S.} \emph{et~al.}
\newblock \bibinfo{title}{{Nonmonotonicity in quantum-to-classical transition
  in multiparticle interference}}.
\newblock \emph{\bibinfo{journal}{Proceedings of the National Academy of
  Sciences}} \textbf{\bibinfo{volume}{110}}, \bibinfo{pages}{1227--1231}
  (\bibinfo{year}{2013}).

\bibitem{Tichy2010}
\bibinfo{author}{Tichy, M.~C.}, \bibinfo{author}{Tiersch, M.},
  \bibinfo{author}{{De Melo}, F.}, \bibinfo{author}{Mintert, F.} \&
  \bibinfo{author}{Buchleitner, A.}
\newblock \bibinfo{title}{{Zero-transmission law for multiport beam
  splitters}}.
\newblock \emph{\bibinfo{journal}{Physical Review Letters}}
  \textbf{\bibinfo{volume}{104}}, \bibinfo{pages}{1--4} (\bibinfo{year}{2010}).

\bibitem{Tichy2013a}
\bibinfo{author}{Tichy, M.}, \bibinfo{author}{Mayer, K.},
  \bibinfo{author}{Buchleitner, A.} \& \bibinfo{author}{M{\o}lmer, K.}
\newblock \bibinfo{title}{{Stringent and efficient assessment of Boson-Sampling
  devices}}.
\newblock \emph{\bibinfo{journal}{Physical Review Letters}}
  \textbf{\bibinfo{volume}{113}}, \bibinfo{pages}{020502}
  (\bibinfo{year}{2014}).

\bibitem{Bonneau2014}
\bibinfo{author}{Bonneau, D.}, \bibinfo{author}{Mendoza, G.~J.},
  \bibinfo{author}{O'Brien, J.~L.} \& \bibinfo{author}{Thompson, M.~G.}
\newblock \bibinfo{title}{{Effect of loss on multiplexed single-photon
  sources}}.
\newblock \emph{\bibinfo{journal}{New Journal of Physics}}
  \textbf{\bibinfo{volume}{17}}, \bibinfo{pages}{1--15} (\bibinfo{year}{2015}).

\bibitem{Shapiro2007}
\bibinfo{author}{Shapiro, J.~H.} \& \bibinfo{author}{Wong, F.~N.}
\newblock \bibinfo{title}{{On-demand single-photon generation using a modular
  array of parametric downconverters with electro-optic polarization
  controls}}.
\newblock \emph{\bibinfo{journal}{Optics Letters}}
  \textbf{\bibinfo{volume}{32}}, \bibinfo{pages}{2698} (\bibinfo{year}{2007}).

\bibitem{Gisin1991}
\bibinfo{author}{Gisin, N.} \& \bibinfo{author}{Thew, R.}
\newblock \bibinfo{title}{{Quantum communication}}.
\newblock \emph{\bibinfo{journal}{Nature photonics}}
  \textbf{\bibinfo{volume}{1}}, \bibinfo{pages}{165--171}
  (\bibinfo{year}{2007}).

\bibitem{Aspuru-Guzik2012}
\bibinfo{author}{Aspuru-Guzik, A.} \& \bibinfo{author}{Walther, P.}
\newblock \bibinfo{title}{{Photonic quantum simulators}}.
\newblock \emph{\bibinfo{journal}{Nature Physics}}
  \textbf{\bibinfo{volume}{8}}, \bibinfo{pages}{285--291}
  (\bibinfo{year}{2012}).

\bibitem{Aaronson2013}
\bibinfo{author}{Aaronson, S.} \& \bibinfo{author}{{attributed to W.S.
  Kolthammer}}.
\newblock \bibinfo{title}{{Scattershot BosonSampling: A new approach to
  scalable BosonSampling experiments.}} (\bibinfo{year}{2013}).
\newblock \urlprefix\url{http://www.scottaaronson.com/blog/?p=1579}.

\bibitem{Lund2013}
\bibinfo{author}{Lund, A.~P.} \emph{et~al.}
\newblock \bibinfo{title}{{Boson Sampling from a Gaussian State}}.
\newblock \emph{\bibinfo{journal}{Phys. Rev. Lett.}}
  \textbf{\bibinfo{volume}{113}}, \bibinfo{pages}{100502}
  (\bibinfo{year}{2014}).

\bibitem{Bentivegna2015}
\bibinfo{author}{Bentivegna, M.} \emph{et~al.}
\newblock \bibinfo{title}{{Experimental scattershot boson sampling}}.
\newblock \emph{\bibinfo{journal}{Science Advances}}
  \textbf{\bibinfo{volume}{1}}, \bibinfo{pages}{1--7} (\bibinfo{year}{2015}).

\bibitem{Lin2007ppg}
\bibinfo{author}{Lin, Q.}, \bibinfo{author}{Yaman, F.} \&
  \bibinfo{author}{Agrawal, G.}
\newblock \bibinfo{title}{{Photon-pair generation in optical fibers through
  four-wave mixing: Role of Raman scattering and pump polarization}}.
\newblock \emph{\bibinfo{journal}{Physical Review A}}
  \textbf{\bibinfo{volume}{75}}, \bibinfo{pages}{23803} (\bibinfo{year}{2007}).

\bibitem{Garay-Palmett2007pps}
\bibinfo{author}{Garay-Palmett, K.} \emph{et~al.}
\newblock \bibinfo{title}{{Photon pair-state preparation with tailored spectral
  properties by spontaneous four-wave mixing in photonic-crystal fiber}}.
\newblock \emph{\bibinfo{journal}{Optics Express}}
  \textbf{\bibinfo{volume}{15}}, \bibinfo{pages}{14870} (\bibinfo{year}{2007}).

\bibitem{Grice2001efs}
\bibinfo{author}{Grice, W.~P.}, \bibinfo{author}{U'Ren, A.~B.} \&
  \bibinfo{author}{Walmsley, I.~A.}
\newblock \bibinfo{title}{{Eliminating frequency and space-time correlations in
  multiphoton states}}.
\newblock \emph{\bibinfo{journal}{Physical Review A}}
  \textbf{\bibinfo{volume}{64}}, \bibinfo{pages}{63815} (\bibinfo{year}{2001}).

\bibitem{Tichy2015spd}
\bibinfo{author}{Tichy, M.~C.}
\newblock \bibinfo{title}{{Sampling of partially distinguishable bosons and the
  relation to the multidimensional permanent}}.
\newblock \emph{\bibinfo{journal}{Physical Review A}}
  \textbf{\bibinfo{volume}{91}} (\bibinfo{year}{2015}).

\bibitem{Aaronson2010ccl}
\bibinfo{author}{Aaronson, S.} \& \bibinfo{author}{Arkhipov, A.}
\newblock \bibinfo{title}{{The Computational Complexity of Linear Optics}}.
\newblock \emph{\bibinfo{journal}{arXiv}}  (\bibinfo{year}{2010}).

\end{thebibliography}

\begin{thebibliography}{10}
\expandafter\ifx\csname url\endcsname\relax
  \def\url#1{\texttt{#1}}\fi
\expandafter\ifx\csname urlprefix\endcsname\relax\def\urlprefix{URL }\fi
\providecommand{\bibinfo}[2]{#2}
\providecommand{\eprint}[2][]{\url{#2}}

\bibitem{Adikan2008}
\bibinfo{author}{Adikan, F. R.~M.}, \bibinfo{author}{Gates, J.~C.},
  \bibinfo{author}{Gawith, C. B.~E.}, \bibinfo{author}{Smith, P. G.~R.} \&
  \bibinfo{author}{Jaque, D.}
\newblock \bibinfo{title}{{Confocal Luminescence Investigations of Two-Beam
  Direct-UV-Written Silica-On-Silicon Waveguides}}.
\newblock \emph{\bibinfo{journal}{IEEE Journal of Quantum Electronics}}
  \textbf{\bibinfo{volume}{44}}, \bibinfo{pages}{1219--1224}
  (\bibinfo{year}{2008}).

\bibitem{Vaccaro2008lfn}
\bibinfo{author}{Vaccaro, L.}, \bibinfo{author}{Cannas, M.} \&
  \bibinfo{author}{Boscaino, R.}
\newblock \bibinfo{title}{{Luminescence features of nonbridging oxygen hole
  centres in silica probed by site-selective excitation with tunable laser}}.
\newblock \emph{\bibinfo{journal}{Solid State Communications}}
  \textbf{\bibinfo{volume}{146}}, \bibinfo{pages}{148--151}
  (\bibinfo{year}{2008}).

\bibitem{Dianov1996ual}
\bibinfo{author}{Dianov, E.~M.}, \bibinfo{author}{Starodubov, D.~S.} \&
  \bibinfo{author}{Frolov, A.~A.}
\newblock \bibinfo{title}{{UV argon laser induced luminescence changes in
  germanosilicate fibre preforms}}.
\newblock \emph{\bibinfo{journal}{Electronics Letters}}
  \textbf{\bibinfo{volume}{32}}, \bibinfo{pages}{246--247}
  (\bibinfo{year}{1996}).

\bibitem{Metcalf2014qtp}
\bibinfo{author}{Metcalf, B.~J.} \emph{et~al.}
\newblock \bibinfo{title}{{Quantum teleportation on a photonic chip}}.
\newblock \emph{\bibinfo{journal}{Nature Photonics}}
  \textbf{\bibinfo{volume}{8}}, \bibinfo{pages}{770--774}
  (\bibinfo{year}{2014}).

\bibitem{Spring2013bsp}
\bibinfo{author}{Spring, J.~B.} \emph{et~al.}
\newblock \bibinfo{title}{{Boson Sampling on a Photonic Chip}}.
\newblock \emph{\bibinfo{journal}{Science (New York, N.Y.)}}
  \textbf{\bibinfo{volume}{339}}, \bibinfo{pages}{798--801}
  (\bibinfo{year}{2013}).

\bibitem{Tichy2015spd}
\bibinfo{author}{Tichy, M.~C.}
\newblock \bibinfo{title}{{Sampling of partially distinguishable bosons and the
  relation to the multidimensional permanent}}.
\newblock \emph{\bibinfo{journal}{Physical Review A}}
  \textbf{\bibinfo{volume}{91}} (\bibinfo{year}{2015}).

\bibitem{Bonneau2014}
\bibinfo{author}{Bonneau, D.}, \bibinfo{author}{Mendoza, G.~J.},
  \bibinfo{author}{O'Brien, J.~L.} \& \bibinfo{author}{Thompson, M.~G.}
\newblock \bibinfo{title}{{Effect of loss on multiplexed single-photon
  sources}}.
\newblock \emph{\bibinfo{journal}{New Journal of Physics}}
  \textbf{\bibinfo{volume}{17}}, \bibinfo{pages}{1--15} (\bibinfo{year}{2015}).

\bibitem{Note1}
\bibinfo{note}{This is only true in the case of 2 or 3 mode devices - for
  unitary devices with four or more modes, there is an additional \protect
  \emph {internal} phase freedom that means there are balanced 4-port devices,
  which do \protect \emph {not} implement the discrete Fourier transform~\cite
  {Bernstein1974}}.

\bibitem{Mauerer2009hci}
\bibinfo{author}{Mauerer, W.}, \bibinfo{author}{Avenhaus, M.},
  \bibinfo{author}{Helwig, W.} \& \bibinfo{author}{Silberhorn, C.}
\newblock \bibinfo{title}{{How colors influence numbers: Photon statistics of
  parametric down-conversion}}.
\newblock \emph{\bibinfo{journal}{Physical Review A}}
  \textbf{\bibinfo{volume}{80}}, \bibinfo{pages}{53815} (\bibinfo{year}{2009}).

\bibitem{Spring2013ocl}
\bibinfo{author}{Spring, J.~B.} \emph{et~al.}
\newblock \bibinfo{title}{{On-chip low loss heralded source of pure single
  photons}}.
\newblock \emph{\bibinfo{journal}{Optics Express}}
  \textbf{\bibinfo{volume}{21}}, \bibinfo{pages}{5932--5935}
  (\bibinfo{year}{2013}).

\bibitem{Spring2014}
\bibinfo{author}{Spring, J.~B.}
\newblock \bibinfo{title}{{Single Photon Generation and Quantum Computing with
  Integrated Photonics}}  (\bibinfo{year}{2014}).

\bibitem{Klyshko1980}
\bibinfo{author}{Klyshko, D.~N.}
\newblock \bibinfo{title}{{Use of two-photon light for absolute calibration of
  photoelectric detectors}}.
\newblock \emph{\bibinfo{journal}{Soviet Journal of Quantum Electronics}}
  \textbf{\bibinfo{volume}{10}}, \bibinfo{pages}{1112} (\bibinfo{year}{1980}).

\bibitem{Eckstein}
\bibinfo{author}{Eckstein, A.}
\newblock \bibinfo{title}{{Mastering quantum light pulses with nonlinear
  waveguide interactions Kontrolle {\"{u}}ber Quantenlichtpulse durch
  nichtlineare Interaktion in Wellenleitern}}.
\newblock \emph{\bibinfo{journal}{Thesis}} .

\bibitem{Bernstein1974}
\bibinfo{author}{Bernstein, H.~J.}
\newblock \bibinfo{title}{{Must quantum theory assume unrestricted
  superposition?}}
\newblock \emph{\bibinfo{journal}{Journal of Mathematical Physics}}
  \textbf{\bibinfo{volume}{15}}, \bibinfo{pages}{1677} (\bibinfo{year}{1974}).

\end{thebibliography}
\end{document}